\documentclass[paper]{JHEP3}
\usepackage{epsfig}
\usepackage{amsmath}
\usepackage{amssymb}
\usepackage{amsbsy}
\usepackage{bbm}

\newcommand{\tmmathbf}[1]{\ensuremath{{\bf #1}}}
\newcommand{\tmop}[1]{\ensuremath{{\rm #1}}}

\def\beq{\begin{equation}}
\def\beqn{\begin{eqnarray}}
\def\eeq{\end{equation}}
\def\eeqn{\end{eqnarray}}

\newcommand\FKS{Frixione, Kunszt and Signer}

\newcommand\HERWIG{{\tt HERWIG}}

\newcommand\PYTHIA{{\tt PYTHIA}}

\newcommand\SISCONE{{\tt SISCONE}}
\newcommand\FASTKT{{\tt KT}}

\newcommand\MINT{{\tt MINT}}

\def\lq{\left[} 
\def\rq{\right]} 
\def\rg{\right\}} 
\def\lg{\left\{} 
\def\({\left(} 
\def\){\right)}

\def\th{\theta}

\newcommand\sss{\mathchoice%
{\displaystyle}%
{\scriptstyle}%
{\scriptscriptstyle}%
{\scriptscriptstyle}%
}

\newcommand\nplus{\oplus}
\newcommand\nminus{\ominus}

\newcommand\splus{{\sss \nplus}}
\newcommand\sminus{{\sss \nminus}}

\newcommand\splusminus{{\mathchoice%
{\vplusminus\displaystyle}%
{\vplusminus\scriptstyle}%
{\vplusminus\scriptscriptstyle}%
{\vplusminus\scriptscriptstyle}%
}}

\newdimen\hbigcirc
\newdimen\wbigcirc

\newdimen\figwidth

\newcommand\captskip{\vskip -0.7cm}

\figwidth=\textwidth

\newcommand\vplusminus[1]{%
\settoheight{\hbigcirc}{$#1\bigcirc$}%
\settowidth{\wbigcirc}{$#1\bigcirc$}%
\makebox[\wbigcirc]{%
\makebox[0pt]{\rule[0.4\hbigcirc]{0.5\wbigcirc}{0.05\hbigcirc}}%
\makebox[0pt]{\rule[0.1\hbigcirc]{0.5\wbigcirc}{0.05\hbigcirc}}%
\makebox[0pt]{\rule[0.1\hbigcirc]{0.05\wbigcirc}{0.6\hbigcirc}}%
\makebox[0pt]{$#1\bigcirc$}}%
}

\newcommand\xplus{x_\splus}
\newcommand\xminus{x_\sminus}
\newcommand\xplusminus{x_\splusminus}
\newcommand\kplus{k_\splus}
\newcommand\kminus{k_\sminus}
\newcommand\Kplus{K_\splus}
\newcommand\Kminus{K_\sminus}
\newcommand\bxplus{\bar{x}_\splus}
\newcommand\bxminus{\bar{x}_\sminus}
\newcommand\bxplusminus{\bar{x}_\splusminus}

\newcommand\as{\alpha_{\sss\rm S}}

\newcommand\Lum{{\cal L}}

\newcommand\pt{p_{\sss\rm T}}
\newcommand\ptmin{{\pt^{\min}}}
\newcommand\kt{k_{\sss\rm T}}

\newcommand\ximax{{\xi_{\sss\rm{M}}}}
\newcommand\boost{\mathbb B}
\newcommand\mur{\mu_{\sss\rm R}}
\newcommand\muf{\mu_{\sss\rm F}}
\newcommand\xitilde{\tilde{\xi}}

\newcommand\matR{{\cal R}}

\newcommand\matRS{\hat{\matR}}

\newcommand\stepf{\theta}

\newcommand\MCatNLO{{\tt MC@NLO}}

\newcommand\xicut{\xi_{c}}

\newcommand\MSB{{\rm \overline{MS}}}

\newcommand\CA{C_{\sss\rm A}}

\newcommand\CF{C_{\sss\rm F}}
\newcommand\TF{T_{\sss\rm F}}

\newcommand\NF{n_{\rm f}}

\newcommand\POWHEG{{\tt POWHEG}}

\newcommand\Rad{\Phi_{\rm rad}}

\newcommand\muF{\mu_{\sss\rm F}}

\newcommand\qb{\bar{q}}

\newcount\minutes 
\newcount\scratch 
 
\def\timestamp{%
\scratch=\time 
\divide\scratch by 60 
\edef\hours{\the\scratch} 
\multiply\scratch by 60 
\minutes=\time 
\advance\minutes by -\scratch 
---$\,$\hours:\null 
\ifnum\minutes< 10 0\fi 
\the\minutes}

\title{NLO Higgs boson production via gluon fusion\\
 matched with shower in {\tt\bf POWHEG}}
\author{Simone Alioli\\
  Universit\`a di Milano-Bicocca and INFN, Sezione di Milano-Bicocca\\
  Piazza della Scienza 3, 20126 Milan, Italy\\
  E-mail: \email{Simone.Alioli@mib.infn.it}}
\author{Paolo Nason\\
  INFN, Sezione di Milano-Bicocca,
  Piazza della Scienza 3, 20126 Milan, Italy\\
  E-mail: \email{Paolo.Nason@mib.infn.it}}
\author{Carlo Oleari\\
  Universit\`a di Milano-Bicocca and INFN, Sezione di Milano-Bicocca\\
  Piazza della Scienza 3, 20126 Milan, Italy\\
  E-mail: \email{Carlo.Oleari@mib.infn.it}}
\author{Emanuele Re\\
  Universit\`a di Milano-Bicocca and INFN, Sezione di Milano-Bicocca\\
  Piazza della Scienza 3, 20126 Milan, Italy\\
  E-mail: \email{Emanuele.Re@mib.infn.it}}
\abstract{
  We present a next-to-leading order calculation of Higgs boson production
  via gluon fusion interfaced to shower Monte Carlo programs,
  implemented according to
  the \POWHEG{} method. A detailed comparison
  with \MCatNLO{} and \PYTHIA{} is carried out for several observables,
  for the Tevatron and LHC colliders.
  Comparisons with next-to-next-to-leading order results and with
  resummed ones are also presented.
}

\begin{document}

\section{Introduction}
Gluon fusion is the
dominant Higgs boson production mechanism both at the Tevatron and at the
LHC. Radiative corrections to this process are known to be
large~\cite{Dawson:1990zj,Djouadi:1991tka,Spira:1995rr}, and it is thus
important that shower
generators that do include them are made available to the experimental
collaborations. In fact, one such generator already exists,
namely the \MCatNLO{} implementation~\cite{Frixione:2002ik} of Higgs boson
production.

In this work we present a next-to-leading order (NLO) calculation of Higgs
boson production via gluon fusion, interfaced to shower Monte Carlo programs
according to the \POWHEG{} method. Unlike the \MCatNLO{} implementation,
our generator produces events with positive (constant) weight, and,
furthermore,
is not tied to the \HERWIG{} shower Monte Carlo program. It can be easily
interfaced to any modern shower generator and, in fact, we show results
of \POWHEG{} interfaced to \HERWIG{}~\cite{Corcella:2000bw,Corcella:2002jc}
and to \PYTHIA{}~\cite{Sjostrand:2006za}.

The \POWHEG{} method was first suggested in ref.~\cite{Nason:2004rx}.
In ref.~\cite{Frixione:2007vw} a detailed general description of its
application to collider processes was given.
Until now, the \POWHEG{} method has been applied to $ZZ$ pair
hadroproduction~\cite{Nason:2006hfa}, heavy-flavour
production~\cite{Frixione:2007nw}, $e^+ e^-$ annihilation into
hadrons~\cite{LatundeDada:2006gx} and into top pairs~\cite{LatundeDada:2008bv},
and Drell-Yan vector boson production~\cite{Alioli:2008gx,Hamilton:2008pd}.
We have built our implementation of the Higgs boson production
by following closely the formulae and results of ref.~\cite{Frixione:2007vw}.

Much of our phenomenological section will be devoted to study the
comparison of our result with that of \MCatNLO. We find fair agreement
between \MCatNLO{} and \POWHEG{} results, except
for the $\pt$ distribution of the Higgs boson, and consequently of the
hardest jet, in the high-$\pt$ region. In this region, the \POWHEG{}
distributions are generally harder.  We have shown that this is due to
next-to-next-to-leading order (NNLO)
effects in the \POWHEG{} formula for the differential cross section.  We
checked that these effects actually bring our result closer to the NNLO
one~\cite{Catani:2007vq}. Other relevant discrepancies are found in the
rapidity difference of the Higgs boson and the 
hardest jet. The dip produced by the
\MCatNLO{} program, found in previous
implementations~\cite{Nason:2006hfa,Frixione:2007nw,Alioli:2008gx}, is
present also here.
We remark that this seems to be a general feature of \MCatNLO{},
since other calculations do not find effects of this
kind~\cite{Mangano:2006rw,Alwall:2007fs,Dittmaier:2008uj}.

The paper is organized as follows. In sec.~\ref{sec:description} we describe
how we performed the calculation for the Higgs boson cross section at the
next-to-leading order. In sec.~\ref{sec:powheg} we discuss the \POWHEG{}
implementation.  In sec.~\ref{sec:results} we show our results for several
kinematic variables and compare them with the \MCatNLO~\cite{Frixione:2002ik}
and \PYTHIA{}~6.4~\cite{Sjostrand:2006za} shower Monte Carlo programs. A
comparison with next-to-next-to-leading order results, as well as with
analytical resummed ones is also carried out.
In sec.~\ref{sec:conc}, we give our conclusions.

\section{Description of the calculation}\label{sec:description}
In this section we fix our kinematic notation, and give the Higgs boson
production differential cross sections up to next-to-leading order in
the strong coupling $\as$.

\subsection{Kinematics}
\subsubsection{Born kinematics}
The Born process has a single partonic contribution, $gg\to H$.  Following
the notation of 
ref.~\cite{Frixione:2007vw}, we denote with $\bar{k}_{\splus}$ and
$\bar{k}_{\sminus}$ the incoming gluon momenta, aligned along the plus and
minus direction of the $z$ axis, and by $\bar{k}_1$ the outgoing Higgs boson
momentum.  If $K_\splus$ and $K_\sminus$ are the momenta of the incoming
hadrons, then we have
\begin{equation}
\bar{k}_\splusminus=\bar{x}_\splusminus K_\splusminus \,,
\end{equation}
where $\bar{x}_\splusminus$ are the momentum fractions, and momentum
conservation reads
\beq
\bar{k}_\splus + \bar{k}_\sminus = \bar{k}_1\,.
\eeq
We introduce the Higgs boson invariant mass squared and rapidity
\begin{equation}
  M^2 = \bar{k}_1^2,\qquad\quad
 Y = \frac{1}{2} \log \frac{\bar{k}_1^0 +\bar{k}_1^3}{\bar{k}_1^0
   -\bar{k}_1^3}\,,  
\end{equation}
so that the set of variables $ \tmmathbf{\bar{\Phi}}_1 \equiv\lg M^2,Y\rg$
fully parametrizes the Born kinematics. From them, we can reconstruct the
momentum fractions
\begin{equation}
\label{eq:xpxm_bar}  
  \bar{x}_{\splus} = \sqrt{\frac{M^2}{S}} e^Y, \qquad\quad
  \bar{x}_{\sminus} = \sqrt{\frac{M^2}{S}}
  e^{- Y},
\end{equation}
where $S=(\Kplus + \Kminus)^2$ is the squared center-of-mass energy of the
hadronic collider.
The Born phase space, in terms of these variables, can be written as
\begin{equation}
  d \tmmathbf{\bar{\Phi}}_1 = d\bar{x}_\splus \, d\bar{x}_\sminus (2\pi)^4 \delta^4(\bar{k}_\splus+
\bar{k}_\sminus -\bar{k}_1) \frac{d^3 \bar{k}_1}{(2\pi)^3 2 \bar{k}_1^0}\,
=\frac{2 \pi}{S} \delta\!\(M^2 -m_H^2\) d M^2 \, d Y \,.
\end{equation}
We generate the Higgs boson virtuality according to a Breit-Wigner
distribution, i.e.~we make the replacement\footnote{In order to compare
our result with other programs, we have also used
slightly different forms of the Breit-Wigner
distribution, that will be illustrated in due time.}
\begin{equation}
\label{eq:breitwigner}
\delta\!\(M^2 -m_H^2\) \, \to \, \frac{1}{\pi} \frac{M^2 \,\Gamma_H
  / m_H}{\(M^2 - m_H^2\)^2 + ( M^2\, \Gamma_H / m_H)^2 }.
\end{equation} 
The decay of the Higgs boson is left to the shower Monte Carlo program, since,
being the Higgs boson a scalar, no spin correlation can arise.

\subsubsection{Real-emission kinematics}
The real emission processes have an additional final-state parton, so
that momentum conservation reads
\begin{equation}
k_\splus + k_\sminus=k_1 + k_2\,,
\end{equation} 
where $k_1$ is the Higgs boson momentum and $k_2$ is the momentum of the
additional final-state parton in the laboratory frame and
\begin{equation}
k_\splusminus=x_\splusminus K_\splusminus \,.
\end{equation}
Since we regularize the infrared divergences in the \FKS{} (FKS) subtraction
scheme~\cite{Frixione:1995ms,Frixione:1997np}, we introduce the appropriate
set of radiation 
variables. In the partonic center-of-mass frame, the final-state parton has
momentum
\begin{equation}
\label{eq:k2}
  k'_2= k_2^{\prime\, 0}\ (1, \sin\theta \sin\phi, \sin\theta \cos\phi,
  \cos\theta), 
\end{equation}
and we use the set $\Rad \equiv\lg\xi, y, \phi\rg$ as radiation variables,
where 
\beq
\label{eq:radvar}
k_2^{\prime\, 0} = \frac{\sqrt{s}}{2} \xi, \qquad y = \cos\th\,,
\eeq
and 
\beq
s=\(k_\splus+k_\sminus\)^2=\frac{M^2}{1-\xi}
\end{equation}
is the partonic center-of-mass energy squared. Since there are no
final-state coloured partons at the Born level, we have to deal with initial-state
singularities only.  
 The soft singularity is characterized by  
$\xi \to 0$, while the collinear limits ($k_2$ parallel to
the $\splus$ or $\sminus$ incoming directions) are characterized by $y \to 1$
and $y \to -1$ respectively. 

\subsubsection{Inverse construction}
The set of variables $\tmmathbf{\Phi}_2\equiv\lg M^2, Y, \xi, y,\phi\rg$ 
fully specifies the real-emission kinematics.
In fact, given these variables, we can reconstruct all the momenta.
Using eq.~(\ref{eq:xpxm_bar}), we can compute the underlying Born momentum
fractions  $\bxplusminus$ and, following sec.~5 of
ref~\cite{Frixione:2007vw}, we have
\begin{equation}
\label{eq:xp_xm_IS_FKS}
 x_\splus = \frac{\bar{x}_\splus}{ \sqrt{1-\xi}} \sqrt{\frac{2-\xi
      (1-y)}{2-\xi (1+y)}}, \qquad
 x_\sminus= \frac{\bar{x}_\sminus}{ \sqrt{1-\xi}} \sqrt{\frac{2-\xi
      (1+y)}{2-\xi(1-y)}},
\end{equation} 
with the kinematics constraints 
\begin{equation}
0 \le \xi \le \ximax(y) \ ,
\end{equation}
where 
\begin{eqnarray}
\label{eq:ximax}
 \ximax(y)= 1- {\rm max}\!\!\!&&\lg
\frac{2(1+y)\,\bar{x}_\splus^2}{\sqrt{(1+\bar{x}_\splus^2)^2(1-y)^2 +
 16\,y\,\bar{x}_\splus^2}+(1-y)(1-\bar{x}_\splus^2)},\right.
\nonumber\\
&&\phantom{\Bigg\{}\left.
\frac{2(1-y)\,\bar{x}_\sminus ^2}{\sqrt{(1+\bar{x}_\sminus ^2)^2(1+y)^2 -
    16\,y\,\bar{x}_\sminus ^2} +(1+y)(1-\bar{x}_\sminus ^2)}\rg .
\end{eqnarray}
The momentum of the final-state parton in the partonic center-of-mass frame
is given by eqs.~(\ref{eq:k2}) and~(\ref{eq:radvar}).  
We then make a longitudinal boost $\boost_L$ from the center-of-mass frame back to the
laboratory frame, with boost velocity
\beq
\beta = \frac{x_\splus - x_\sminus}{x_\splus + x_\sminus}\,,
\eeq
to obtain $k_2$ from $k_2'$
\beq
k_2 = \boost_L \, k_2'\,.
\eeq
From momentum conservation, we reconstruct the Higgs boson momentum
\beq
k_{1}=\xplus \Kplus+\xminus \Kminus -k_{2}\,.
\eeq
Finally, the two-body phase space can  be written in a factorized form in
terms of the Born and radiation phase space
\beq
  d \tmmathbf{\Phi}_2 = dx_\splus \, dx_\sminus (2\pi)^4 \delta^4
  (k_\splus+ k_\sminus -k_1-k_2) \frac{d^3 k_1}{(2\pi)^3 2 k_1^0}
  \frac{d^3 k_2}{(2\pi)^3 2 k_2^0}
= d\tmmathbf{\bar{\Phi}}_1\,d\Rad  \,,
\eeq
where
\begin{equation}
\label{eq:dRad_IS_FKS}
d\Rad=\frac{M^2}{(4\pi)^3}\,\frac{\xi}{\(1-\xi\)^2}\,d\xi\,dy\, d\phi\,.
\end{equation}

\subsection{Cross sections}
In order to apply the \POWHEG{} method, we need the Born, real and virtual
contributions to the differential cross section, i.e.~the squared amplitudes,
averaged over colours and helicities of the incoming partons, and multiplied
by the appropriate flux factor.

\subsubsection{Born contribution}
At Born level, Higgs boson production via gluon fusion proceeds through the
coupling of the Higgs boson to a heavy-quark loop.
The squared matrix element for the lowest-order contribution, averaged over
colours and helicities of the incoming gluons, 
and multiplied by the flux factor $1/(2M^2)$, is given by
\begin{equation}
\label{eq:Bgg}
\mathcal{B}_{g g} = \frac{\as^2}{\pi^2} \frac{G_F\, M^2}{576\,
\sqrt{2}}
\,\Bigg| \,\frac{3}{2}\sum_Q \tau_Q \Big[ 1+ (1-\tau_Q)
f(\tau_Q)\Big]
\Bigg|^2 \,,
\end{equation}
where
$\tau_Q = 4 m^2_Q / M^2$, and the sum runs over the heavy flavours with mass
$m_Q$ circulating in the loop.  The function $f$ is given by
\begin{eqnarray}
f(\tau_Q) &=& \lg 
\begin{array}{ll} 
{\displaystyle \arcsin^2 \frac{1}{\sqrt{\tau_Q}}} &   \tau_Q  \geq 1 \,, 
\\[4mm]
{\displaystyle -\frac{1}{4} \lq \log \( \frac{ 1 + \sqrt{(1-\tau_Q)}}{ 1 -
  \sqrt{(1-\tau_Q)}}\) - i \pi\rq^2 }  \qquad &  \tau_Q < 1 .
\end{array} \right.
\end{eqnarray}
In our implementation we only retain the contribution coming from the top quark.
 
\subsubsection{Virtual corrections}
In the calculation of all NLO corrections, we have used an
effective Lagrangian, where the heavy-quark degrees of freedom have been
integrated out. This corresponds to take the $m_Q \to \infty$ limit.

We have regularized the infrared divergences according to the conventional
dimensional regularization method, i.e.~we have set the space-time dimensions
$D=4-2\epsilon$.

The finite soft-virtual term, obtained from the sum of the divergent virtual
contributions and of the integral over the radiation variables of the
counter-terms is given by (see eq.~(2.99) of ref.~{\cite{Frixione:2007vw}})
\begin{equation}
  \mathcal{V}_{g g } = \frac{\as}{2\pi} \lq 
- \(\frac{11}{3}\, \CA - \frac{4}{3}\, \TF n_f \) \log\frac{\muf^2}{\mur^2}
+ \frac{11}{3}\CA + \frac{2\pi^2}{3}\CA  \rq \mathcal{B}_{gg}\,.
\end{equation}
In deriving this equation we have set $\xicut=1$.
We indicate with $\mur$ and $\muf$ the renormalization and factorization
scales, respectively.

\subsubsection{Real corrections}

At NLO, there are four subprocesses that contribute to Higgs boson
production: $gg\to Hg$,  $gq\to Hq$, $qg\to Hq$ and  $q\qb\to Hg$, where 
$q$ runs over all possible quark and antiquark flavours and $q$ and $\qb$ are
conjugate in flavour.
The respective squared amplitudes, averaged over the incoming helicities and
colours and  multiplied by the flux factor $1/(2s)$ are given by 
\beqn
\matR_{gg} &=& \frac{\as^3}{12 \pi} \,\frac{G_F}{\sqrt{2}} \,\frac{1}{2s}\,
\frac{s^4+t^4+u^4+M^8}{s t u}\,,\\
\matR_{gq} &=& - \frac{\as^3}{27 \pi} \,\frac{G_F}{\sqrt{2}}\, \frac{1}{2s}
\,\frac{s^2+u^2}{t}\,,\\
\matR_{qg} &=& -\frac{\as^3}{27 \pi}\, \frac{G_F}{\sqrt{2}}\, \frac{1}{2s}\,
\frac{s^2+t^2}{u}\,,\\
\matR_{q\qb} &=& \frac{ 8 \,\as^3}{81  \pi}\, \frac{G_F}{\sqrt{2}}
\,\frac{1}{2s} \,\frac{t^2+u^2}{s}\,,
\eeqn
where
\beq
s= (\kplus+\kminus)^2=\frac{M^2}{1-\xi}, \quad t= (\kminus-k_2)^2=-
\frac{s}{2}\, \xi\, (1+y), \quad u=(\kplus-k_2)^2=- \frac{s}{2}\, \xi\,
(1-y). 
\eeq 
In terms of the FKS variables we then have
\beqn
\matR_{gg} &=& \frac{\as^3}{12 \pi} \,\frac{G_F}{\sqrt{2}}\, \frac{1}{4}
\lq 8+ \( y^4+6 y^2 +1 \) \xi^4 + 8(1-\xi)^4 \rq
\frac{1}{\xi^2 (1-y^2)} \,,
\\ 
\matR_{gq} &=&  \frac{\as^3}{27 \pi} \,\frac{G_F}{\sqrt{2}} \,
\frac{1}{4} \lq  4+(1-y)^2 \xi^2 \rq \frac{1}{\xi (1+y)} \,,
\\
\matR_{qg} &=& \frac{\as^3}{27 \pi}\, \frac{G_F}{\sqrt{2}} \,
\frac{1}{4} \lq   4+(1+y)^2 \xi^2 \rq \frac{1}{\xi (1-y)} \,,
\\
\matR_{q\qb} &=& \frac{ 8 \,\as^3}{81  \pi} \,\frac{G_F}{\sqrt{2}}\,
\frac{1}{4} \lq \xi^2 \(1+y^2\)\rq\,,
\eeqn
where the singular behavior for a soft ($\xi\to 0$) or collinear gluon
($y \to \pm 1$) is clearly manifest. Notice that the contribution
$\matR_{q\qb}$ is not singular and has no underlying Born.

\subsubsection{Collinear remnants}
After the subtraction of the initial-state collinear singularities into the
parton distribution functions, finite collinear remnants are left over.  The
kinematics of these terms is Born-like. More precisely, we can introduce
two sets of variables, $\bar{\bf \Phi}_{1,\splusminus}=\lg M^2,Y,z\rg$, such
that momentum conservation reads 
\beq
z \,x_\splus K_\splus + x_\sminus K_\sminus =
k_1
\eeq
for the $\splus$ direction and 
\beq
x_\splus K_\splus + z\, x_\sminus
K_\sminus = k_1
\eeq
for the $\sminus$ one.  We can then associate an underlying
Born configuration $\bar{\bf \Phi}_{1}$ such that
\beq
\bar{k}_\splus = z \,x_\splus K_\splus\,, \qquad \bar{k}_\sminus = x_\sminus
K_\sminus\,, \qquad \bar{k}_1 = k_1 
\eeq
for the $\splus$ direction, and 
\beq
\bar{k}_\splus = x_\splus K_\splus\,, \qquad \bar{k}_\sminus = z\,x_\sminus
K_\sminus\,, \qquad \bar{k}_1 = k_1 
\eeq
for the $\sminus$ one.

The collinear remnants are given in eq.~(2.102) of
ref.~{\cite{Frixione:2007vw}}, where we have fixed $\xicut=1$ and
$\delta_I=2$ and chosen the $\MSB$ renormalization scheme. For the $\splus$
direction and for the two different real-term contributions, they are given
by
\begin{eqnarray}
\nonumber
\mathcal{G}_{\splus}^{q g}\(\bar{\bf \Phi}_{1,\splus}\) 
&=& \frac{\as}{2\pi} \, \CF \lg (1-z) \frac{1+(1-z)^2}{z}\lq 
\(\frac{1}{1-z}\)_{\!\!+} \log \(\frac{M^2}{z \mu^2_F}\) \right. \right.\\   
&&\hspace{5cm} \left.\left. +\  
2\( \frac{\log(1-z)}{1-z}\)_{\!\!+} \rq +  z \rg {\cal B}_{g g}\,,\\\nonumber
\\\nonumber
\mathcal{G}_{\splus}^{g g}\(\bar{\bf \Phi}_{1,\splus}\) 
\!&=&\! \frac{\as}{2\pi} \,2\CA \lq z + \frac{(1-z)^2}{z} +
z(1-z)^2 \rq\! \lq 
\(\frac{1}{1-z}\)_{\!\!+} \log \(\frac{M^2}{z \mu^2_F}\)   \right.\\   
&&\hspace{6cm} \left. +\  2\(
\frac{\log(1-z)}{1-z}\)_{\!\!+} \rq  {\cal B}_{g g}\,.
\end{eqnarray}
The other two collinear remnants,
$\mathcal{G}_{\sminus}^{g q}(\bar{\bf \Phi}_{1,\sminus})$ and
$\mathcal{G}_{\sminus}^{g g}(\bar{\bf \Phi}_{1,\sminus})$,  have the same
functional form of $\mathcal{G}_{\splus}^{q g}\(\bar{\bf \Phi}_{1,\splus}\)$
and $\mathcal{G}_{\splus}^{g g}\(\bar{\bf \Phi}_{1,\splus}\)$ respectively,
since $ {\cal B}_{g g}$ only depends upon $k_1^2$.

\section{\POWHEG{} implementation}
\label{sec:powheg}

\subsection{Generation of the Born variables}
The first step in the \POWHEG{} implementation is the generation of the Born
kinematics. According to ref.~\cite{Frixione:2007vw}, we introduce the 
$\bar{B}\!\(\bar{\bf \Phi}_1\)$ function, defined as
\begin{eqnarray}
\label{eq:Bbar}
\bar{B}\!\(\bar{\bf \Phi}_1\) \!&=& B_{gg}\!\(\bar{\bf \Phi}_1\) + 
V_{gg}\!\(\bar{\bf \Phi}_1\) + \int d \Rad \bigg\{ \hat{R}_{gg}\(\bar{\bf
  \Phi}_1,\Rad\)  
\nonumber\\
&&  + \sum_q\lq \hat{R}_{qg}\(\bar{\bf \Phi}_1,\Rad\)  +
\hat{R}_{gq}\(\bar{\bf 
  \Phi}_1,\Rad\) \rq \bigg\} 
\nonumber\\
&&+\int_{\bar{x}_{\splus}}^1 \frac{dz}{z}
\lq G_{\splus}^{gg}(\bar{\bf \Phi}_{1,\splus})+\sum_q G_{\splus}^{qg}(\bar{\bf
  \Phi}_{1,\splus})\rq   
+  \int_{\bar{x}_{\sminus}}^1 \frac{dz}{z} \lq G_{\sminus}^{gg}(\bar{\bf
  \Phi}_{1,\sminus})+ 
\sum_q G_{\sminus}^{gq}(\bar{\bf \Phi}_{1,\sminus})  \rq,
\nonumber
\\
\end{eqnarray}
where
\begin{eqnarray}
B_{gg}\!\(\bar{\bf \Phi}_1\) &=& \mathcal{B}_{gg}\!\(\bar{\bf \Phi}_1\) 
\, \Lum_{gg}\!\(\bar{x}_{\splus},\bar{x}_{\sminus}\),\\
V_{gg}\!\(\bar{\bf \Phi}_1\) &=& \mathcal{V}_{gg}\!\(\bar{\bf \Phi}_1\)
\, \Lum_{gg}\!\(\bar{x}_{\splus},\bar{x}_{\sminus}\),\\
\hat{R}_{gg}(\bar{\bf \Phi}_1,\Rad) &=&
  \hat{\mathcal{R}}_{gg}(\bar{\bf \Phi}_1,\Rad)  
\, \Lum_{gg}\!\(x_{\splus},x_{\sminus}\),\\
\hat{R}_{gq}(\bar{\bf \Phi}_1,\Rad)  &=&
\hat{\mathcal{R}}_{gq}(\bar{\bf\Phi}_1,\Rad)  
\, \Lum_{gq}\!\(x_{\splus},x_{\sminus}\),\\
\hat{R}_{qg}(\bar{\bf \Phi}_1,\Rad) &=& \hat{\mathcal{R}}_{qg}(\bar{\bf
  \Phi}_1,\Rad) 
\, \Lum_{qg}\!\(x_{\splus},x_{\sminus}\),\\
%
G_{\splus}^{gg}(\bar{\bf \Phi}_{1,\splus}) &=& 
\mathcal{G}_{\splus}^{gg}(\bar{\bf \Phi}_{1,\splus}) 
\, \Lum_{gg}\!\( \frac{\bar{x}_{\splus}}{z},\bar{x}_{\sminus}\),\\
G_{\splus}^{qg}(\bar{\bf \Phi}_{1,\splus}) &=&
\mathcal{G}_{\splus}^{qg}(\bar{\bf \Phi}_{1,\splus})  
\, \Lum_{qg}\!\( \frac{\bar{x}_{\splus}}{z},\bar{x}_{\sminus}\),\\
G_{\sminus}^{gg}(\bar{\bf \Phi}_{1,\sminus}) &=&
\mathcal{G}_{\sminus}^{gg}(\bar{\bf \Phi}_{1,\sminus})  
\, \Lum_{gg}\!\(\bar{x}_{\splus},\frac{\bar{x}_{\sminus}}{z}\),\\
G_{\sminus}^{gq}(\bar{\bf \Phi}_{1,\sminus}) &=&
\mathcal{G}_{\sminus}^{gq}(\bar{\bf   \Phi}_{1,\sminus})  
\, \Lum_{gq}\!\(\bar{x}_{\splus},\frac{\bar{x}_{\sminus}}{z}\),
\end{eqnarray}
with $x_{\splus}, x_{\sminus}$ given in eq.~(\ref{eq:xp_xm_IS_FKS}) and the
luminosity $\mathcal{L}$ is defined in terms of the parton distribution
functions $f_f^{\splusminus}(\xplusminus, \muF^2)$
\begin{equation}
\label{eq:luminosity}
\Lum_{ff'}\!\(\xplus,\xminus\) =  f_f^{\splus}(\xplus, \muF^2)\;
  f_{f'}^{\sminus} (\xminus, \muF^2)\,.
\end{equation}
Observe that the $\matR_{q\qb}$ term does not appear in $\bar{B}$, since
it does not have a valid underlying Born. It is just generated
separately, as described at the end of this section.

All the integrals appearing in eq.~(\ref{eq:Bbar}) are finite. In fact,
according the the FKS subtraction scheme, the hatted functions
\beq
\label{eq:FKSrin}
\hat{\matR}_{ij} = \frac{1}{\xi} 
\lg \frac{1}{2} \(\frac{1}{\xi}\)_{\!\!+}
\lq\(\frac{1}{1-y}\)_{\!\!+} \!\!+ \(\frac{1}{1+y}\)_{\!\!+} \rq\rg
\lq \(1-y^2\) \, \xi^2 \, \matR_{ij} \rq
\eeq
have only integrable divergences.
Some care should still be used when dealing with the plus distributions.  
In order to illustrate this, we explicitly show how to deal with the
$\matR_{gg}$ term, that is the most singular one.
According to  eq.~(\ref{eq:FKSrin}), it can be written
\beq
\label{eq:Rhat_gg}
\matRS_{gg} = \frac{\as^3}{12 \pi} \,\frac{G_F}{\sqrt{2}}
\lq 2+ \frac{y^4+6 y^2 +1}{4} \xi^4 + 2(1-\xi)^4 \rq
\lg \frac{1}{2} \(\frac{1}{\xi}\)_{\!\!+}
\lq\(\frac{1}{1-y}\)_{\!\!+} \!\!+ \(\frac{1}{1+y}\)_{\!\!+} \rq\rg
\frac{1}{\xi}.
\eeq
Inserting now the expression~(\ref{eq:dRad_IS_FKS}) of $d\Rad$ into
eq.~(\ref{eq:Bbar}), we have
\beq
\int d\Rad \,\Lum_{gg}\!\(\xplus,\xminus\) \matRS_{gg} =
\frac{M^2}{(4\pi)^3} \int_{-1}^{1} dy\int_0^{\ximax(y)}\!\! d\xi
\,\frac{\xi}{\(1-\xi\)^2} \int_0^{2\pi}\!\! d\phi
\;\Lum_{gg}\!\(\xplus,\xminus\) \matRS_{gg}\,,
\end{equation}
where $\ximax(y)$ is given in eq.~(\ref{eq:ximax}).
The integration over the azimuthal angle $\phi$ is straightforward, giving an
overall multiplicative factor of $2\pi$. Considering then the
$(1/(1-y))_+$ term 
only, we get an integral of the form
\beq
I = \int_{-1}^{1} dy\int_0^{\ximax(y)} \!\! d\xi 
\(\frac{1}{\xi}\)_{\!\!+} \(\frac{1}{1-y}\)_{\!\!+} \!f(\xi,y)
\,\Lum_{gg}\!\(\xplus,\xminus\) 
\eeq
where  
\beq
f(\xi,y) = \frac{\as^3}{12 \pi} \,\frac{G_F}{\sqrt{2}}
\lq 2+ \frac{y^4+6 y^2 +1}{4} \xi^4 + 2(1-\xi)^4 \rq \frac{1}{2}
\frac{M^2}{(4\pi)^3}\,\frac{2\pi}{\(1-\xi\)^2}\,.
\eeq
Recalling the definition of the plus distributions
\begin{eqnarray}
\label{eq:uoxidef}
\int_0^1 d\xi \, \(\frac{1}{\xi}\)_{\!\!+} f(\xi) &=& \int_0^1 d\xi\,
\frac{f(\xi)-f(0)}{\xi}\,,
\\
\label{eq:uoyipmdef}
\int_{-1}^{1}dy \, \(\frac{1}{1-y}\)_{\!\!+} f(y) &=& \int_{-1}^{1}dy\,
\frac{f(y)-f(1)}{1- y}\,,
\end{eqnarray}
and making the change of variable
\beq
\xi = \ximax(y) \, \xitilde\,,
\eeq
we are left with
\beqn
I &=&  \int_{-1}^1 dy \(\frac{1}{1-y}\)_{\!\!+} \int_0^1 d\xitilde
 \(\frac{1}{\xitilde}\)_{\!\!+} f\(\xi,y\) \,
\Lum_{gg}\(\xplus,\xminus\)  
\nonumber\\
&&+\int_{-1}^1 dy \(\frac{1}{1-y}\)_{\!\!+} 
 f(0,y) \,\log\ximax(y) \; \Lum_{gg}\!\(\bxplus,\bxminus\) 
\nonumber\\
 &=& \int_{-1}^1 dy \(\frac{1}{1-y}\)_{\!\!+} \int_0^1 d\xitilde
\,\frac{1}{\xitilde}
\Big[  f\(\xi,y\) \Lum_{gg}\!\(\xplus,\xminus\)
- f\(0,y\) \Lum_{gg}\!\(\bxplus,\bxminus\)  \Big]
\nonumber\\
&&+\int_{-1}^1 dy\, \frac{1}{1-y}
\Big[  f(0,y) \,\log\ximax(y) 
-  f(0,1) \,\log\ximax(1)\Big] \Lum_{gg}\!\(\bxplus,\bxminus\) 
\nonumber\\
 &=& \int_{0}^1 d\tilde{y} \int_0^1 d\xitilde\,
\frac{1}{1-\tilde{y}}\, \frac{1}{\xitilde}
\Bigg\{
\Big[  f\(\xi,y\) \Lum_{gg}\!\(\xplus,\xminus\)
- f\(0,y\) \Lum_{gg}\!\(\bxplus,\bxminus\)  \Big]
\nonumber\\
 &&
\hspace{3.45cm}-\lq  f\(\xi,1\) \Lum_{gg}\!\(\frac{\bxplus}{1-\xi},\bxminus\)
- f\(0,1\) \Lum_{gg}\!\(\bxplus,\bxminus\)  \rq
\Bigg\}
\nonumber\\
&&+\int_{0}^1 d\tilde{y}\,\frac{1}{1-\tilde{y}}
\Big[ f(0,y)\, \log\ximax(y) - f(0,1) \,\log\ximax(1) 
\Big] \Lum_{gg}\!\(\bxplus,\bxminus\) ,
\eeqn
where we have used the expression of $\xplusminus$ of
eq.~(\ref{eq:xp_xm_IS_FKS}) and $\ximax(1)=1-\bxplus$ (see
eq.~(\ref{eq:ximax})). In the last line we have made the 
further change of variable
\beq
y= -1 + 2 \,\tilde{y}\,,
\eeq
so that all radiation variables are mapped into a cubic unit volume.
The integral $I$ is now manifestly finite and can be computed numerically.

The same manipulations should be applied to the $z$ integration of the
collinear remnants in eq.~(\ref{eq:Bbar}). For example, concentrating on the
two plus distributions in the $\mathcal{G}_{\splus}^{gg}$ term, we have to
deal with integrals of the form
\begin{eqnarray}
  \int_{\bar{x}_{\splus}}^1 d z \left( \frac{1}{1 - z} \right)_+ f (z) & = & 
  \log (1 - \bar{x}_{\splus}) f (1) + \int_0^1 d \tilde{\xi}\, \frac{f
  (z) - f (1)}{1 - \tilde{\xi}}\,,\\
  \int_{\bar{x}_{\splus}}^1 d z \left( \frac{\log (1 - z)}{1 - z} \right)_+ f
  (z) & = & \frac{1}{2} \log^2 (1 - \bar{x}_{\splus}) f (1) +
  \int_0^1 d \tilde{\xi} \,\frac{\log (1 - z)}{1 - \tilde{\xi}} \lq f (z) - f
  (1)\rq , 
\phantom{aaaa}
\end{eqnarray}
where $f(z)$ is finite in the $z\to 1$ limit and we have made the change of
variable
\beq
z =  \bar{x}_{\splus} + \tilde{\xi} \(1 - \bar{x}_{\splus}\)\,.
\eeq
At the end of this procedure, the most general form one can obtain for
$\bar{B}$  is
\begin{equation}
  \bar{B}\!\(\bar{\bf \Phi}_1\) = D\!\( \bar{\bf \Phi}_1\) + \int_0^1
  d\tilde{\xi}\, E\!\(\bar{\bf \Phi}_1,\tilde{\xi}\) + \int_{0}^1 d \tilde{y}
  \int_0^{1} d \tilde{\xi}\, F\!\(\bar{\bf \Phi}_1,\tilde{\xi},\tilde{y}\),
\end{equation}
and we can define the function
\begin{equation}
  \tilde{B}\!\(\bar{\bf \Phi}_1,\tilde{\xi},\tilde{y}\) = D\!\( \bar{\bf
    \Phi}_1\) + E\!\(\bar{\bf \Phi}_1,\tilde{\xi}\) + 
   F\!\(\bar{\bf \Phi}_1,\tilde{\xi},\tilde{y}\)\,,
\end{equation}
so that 
\begin{equation}
\bar{B}\!\(\bar{\bf \Phi}_1\) = \int_{0}^1 d \tilde{y}
  \int_0^{1} d \tilde{\xi}\, \tilde{B}\!\(\bar{\bf
    \Phi}_1,\tilde{\xi},\tilde{y}\) . 
\end{equation}
In order to generate the underlying Born kinematics,
we first compute the two distinct contributions to the total cross section,
defined by
\begin{equation}
\sigma_{\rm
  tot}=\sigma_{\scriptscriptstyle\bar{B}}+\sum_q\sigma_{\scriptscriptstyle 
  R_{q\qb}}\,, 
\end{equation}
where
\begin{eqnarray}
\sigma_{\scriptscriptstyle\bar{B}}&=&\int d \bar{\bf \Phi}_1 \,
\bar{B}\!\(\bar{\bf \Phi}_1\), 
\nonumber \\
\sigma_{\scriptscriptstyle R_{q\qb}}&=&\int d \bar{\bf \Phi}_1 \, d\Rad\,
R_{q\qb}(\bar{\bf \Phi}_1 ,\Rad) \;,
\end{eqnarray}
and 
\beq 
R_{q\qb}(\bar{\bf \Phi}_1 ,\Rad) = \matR_{q\qb}(\bar{\bf \Phi}_1
,\Rad) \, \Lum_{q\qb}(\xplus,\xminus)\,.  
\eeq 
We then decide whether the event is a $\bar{B}$ event or a $R_{q\qb}$ one,
with a probability equal to $\sigma_{\scriptscriptstyle\bar{B}}/\sigma_{\rm
  tot}$ and $\sigma_{\scriptscriptstyle R_{q\qb}}/\sigma_{\rm tot}$
respectively.  In case of a $\bar{B}$ event, the generation of the Born
variables $\bar{\bf \Phi}_1$ is performed by using the integrator-unweighter
program \MINT{} \cite{Nason:2007vt} that, after a single integration of the
function $\tilde{B} (\bar{\bf \Phi}_1,\tilde{\xi},\tilde{y} )$ over the
Born and radiation variables, can generate a set of values for the variables
$\{ \bar{\bf \Phi}_1,\tilde{\xi},\tilde{y} \}$, distributed according to the
weight $\tilde{B}(\bar{\bf\Phi}_1,\tilde{\xi},\tilde{y})$. We then keep
the $\bar{\bf \Phi}_1$ generated values only, and neglect all the others, which
corresponds to integrate over them. The event is then further processed, to
generate the radiation variables, as illustrated in the following section.
In case of a $R_{q\qb}$ event, one uses the same method used for the
$\bar{B}$ case, except that, at the end, one keeps the whole set of Born plus
radiation variables, that fully defines the kinematics of a real event.  In
this last case, one does not need to do anything else, and the event is
passed to the Les Houches Interface, to be further showered by the Monte
Carlo program.

\subsection{Generation of the radiation variables}
Radiation kinematics is generated using the \POWHEG{} Sudakov form
factor
\begin{eqnarray}
  \label{eq:Vsuda} 
  \Delta\!\(\bar{\bf \Phi}_1, \pt\) &=& \exp
  \lg  - \int  d\Rad \,  \frac{R\(\bar{\bf \Phi}_1,\Rad \) }
{B\!\(\bar{\bf \Phi}_1\)}
  \,\stepf\!\(\kt - \pt\)\! \rg\!, 
\end{eqnarray}
where we have defined
\beqn
R\(\bar{\bf \Phi}_1,\Rad \) &=&
R_{gg}\(\bar{\bf \Phi}_1,\Rad \) 
 + \sum_q \lq R_{qg}\(\bar{\bf \Phi}_1,\Rad\)  + R_{gq}\(\bar{\bf
  \Phi}_1,\Rad\)\rq \,,\\
B\!\(\bar{\bf \Phi}_1\) &=& B_{gg}\!\(\bar{\bf \Phi}_1\)\,, 
\eeqn
and 
\begin{equation}
  \kt^2 = \frac{s}{4}\, \xi^2 \(1 - y^2\) =
 \frac{M^2}{4 (1 - \xi)} \,\xi^2 \(1 -  y^2\)
\end{equation}
is the exact squared transverse momentum of the radiated parton. The
factorization and renormalization scales in eq.~(\ref{eq:Vsuda}) should be
taken equal to $\kt^2$, in order to recover the correct leading logarithm (LL) Sudakov
behavior\footnote{We will show in sec.~\ref{sec:nllresum} how it is possible
  to reach next-to-leading logarithmic accuracy for this particular process.}.
  
To generate the radiation variables, we use the veto method.
This requires to find a simple upper bound for
the integrand in eq.~(\ref{eq:Vsuda})
\begin{equation}
\label{eq:upperb}
  \frac{M^2}{(4 \pi)^3}\, \frac{\xi}{\(1 - \xi\)^2} \,
\frac{
  R\!\(\bar{\bf \Phi}_1,\Rad\)}{B\!\(\bar{\bf \Phi}_1\)}\,.
\end{equation}
A suitable upper bounding function is given by
\begin{equation}
  \label{eq:ubound} 
U = N \,\frac{\as\!\(\kt^2\)}{\xi\(1 - y^2\)}\,,
\end{equation}
where $N$ is determined by spanning randomly the whole phase space and
imposing that $U$ is larger than the integrand function.
The generation of the event according to the bound
(\ref{eq:ubound}) is documented in great detail in Appendix~D of
ref.~{\cite{Nason:2006hf}}, and we do not repeat it here.  \\

The \POWHEG{} differential cross section for the generation of the hardest event
is given by
\begin{eqnarray}
\label{eq:POWHEGsigmasimple}
d\sigma &=& \bar{B}(\bar{\bf \Phi}_1)\, d \bar{\bf \Phi}_1
 \lg \Delta\!\(\bar{\bf \Phi}_1,\ptmin\)
+
\Delta\!\(\bar{\bf \Phi}_1, \pt \)\, \frac{
 R\(\bar{\bf \Phi}_1,\Rad \)}{ B\!\(\bar{\bf \Phi}_1\)} \, d\Rad \rg 
\nonumber\\ 
&& \hspace{6.5cm}  +\ \sum_q R_{q \bar q}\(\bar{\bf
  \Phi}_1,\Rad \) d \bar{\bf \Phi}_1 d\Rad \,,\qquad
\end{eqnarray}
where the last term in the sum is the non-singular real
contribution.  In the $\bar B$ and $R_{q \bar q}$ functions, the
renormalization and factorization scales, $\mur$ and $\muf$, should be taken
of the order of the hard scale of the process, i.e. the Higgs boson mass or
its transverse mass.  During the generation of radiation, the two scales
should instead be taken equal to the transverse momentum of the produced
radiation, in order to recover the correct Sudakov form factor.

We remark that, in the formula for the strong coupling constant
used for the generation of radiation, we have
properly taken into account the heavy-flavour thresholds.  That is to say,
when the renormalization scale $\mur$ crosses a heavy-flavour mass threshold,
we change the number of active flavours accordingly. Furthermore, as
discussed in refs.~\cite{Frixione:2007vw,Nason:2006hfa},
we use a rescaled value $\Lambda_{\scriptscriptstyle\rm{MC}}=
 1.569\,\Lambda_{\scriptscriptstyle\overline{\rm MS}}^{(5)}$
in the expression for $\as$, in order to achieve next-to-leading
logarithmic accuracy in the Sudakov form factor  (see
sec.~\ref{sec:nllresum} for more details).

\section{Results}
\label{sec:results}

In this section we present our results, obtained for the Tevatron and the
LHC, and the comparison done with \MCatNLO{} and \PYTHIA.  We have used the
CTEQ6M~\cite{Pumplin:2002vw} set for the parton distribution functions and the
corresponding returned value $\Lambda_{\scriptscriptstyle\overline{\rm
    MS}}^{(5)}=0.226$~GeV. In the generation of the radiation, we have fixed
the  lower cutoff of the transverse momentum to the value
$\ptmin = \sqrt{5}\,
\Lambda_{\scriptscriptstyle\rm{MC}}$.
The renormalization and factorization scales have
been taken equal to the Higgs boson transverse mass $m^H_{\sss\rm T} =
\sqrt{m^2_H + (\pt^H)^2}$.

No acceptance cuts have been applied in any of the following plots.

\subsection{\POWHEG{} - \MCatNLO\ comparison}
We have compared our results with \MCatNLO, the only existing program where
NLO Higgs boson production via gluon fusion  is merged with a shower Monte
Carlo program.  Since \MCatNLO{} uses only the
\HERWIG~\cite{Corcella:2000bw,Corcella:2002jc} angular-ordered shower, we
have also interfaced \POWHEG{} with \HERWIG, in order to
minimize effects due to differences in the  shower and hadronization
algorithms.

\MCatNLO{} generates the Higgs boson virtuality  $M^2$ according to the
Breit-Wigner form  
\beq
\label{eq:bwsimple} 
\frac{1}{\pi} \frac{m_H \, \Gamma_H} {\(M^2 -
m_H^2\)^2 + ( m_H \, \Gamma_H )^2 } \,.
\eeq 
For the purpose of this comparison we have thus used the same form.
We have considered two
different sets of values for the Higgs boson mass and width: 
$m_H=120$~GeV with $\Gamma_H=3.605$~MeV and 
$m_H=400$~GeV with $\Gamma_H =28.89$~GeV.

Both in \POWHEG{} and in \MCatNLO{} there is the option to retain the full
top-mass dependence in the Born cross section, i.e.~to use a finite $\tau_Q$
value in eq.~(\ref{eq:Bgg}).  We have then the choice to generate our Born
variables by fixing $m_t=171$~GeV in the $\bar{B}$ term in
eq.~(\ref{eq:Bbar}) or by sending $m_t\to\infty$.  Since we have computed the
real-radiation term only in the $m_t\to\infty$ limit, we have to use the same
limit in the calculation of the Born term in the Sudakov form factor~(\ref{eq:Vsuda}), in
order to recover the correct Altarelli-Parisi behavior when the collinear limit
is approached.

\subsubsection{Tevatron results}

\begin{figure}[htb]
\begin{center}
\epsfig{file=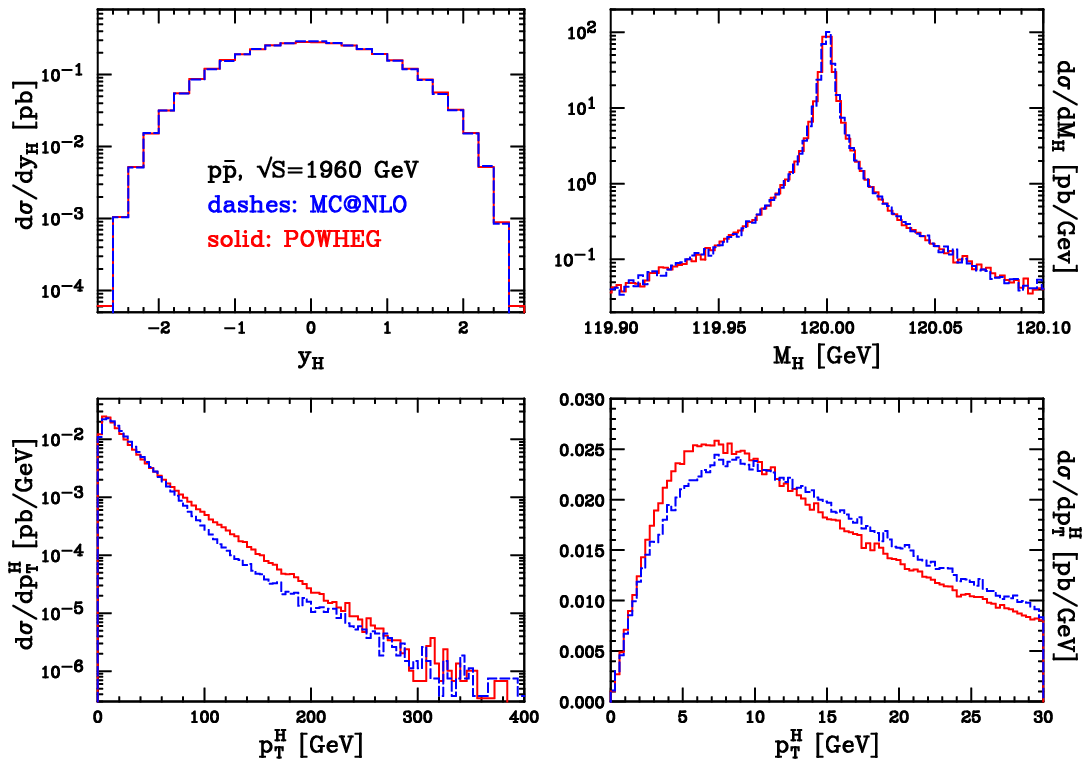,width=\figwidth}
\end{center}
\captskip
\caption{\label{fig:cmp1-tev-mcatnlo}
Comparison between \POWHEG{} and \MCatNLO{} for the rapidity, invariant
mass and transverse-momentum distributions of a Higgs boson with $m_H
=120$~GeV, at Tevatron $p\bar p$ collider.} 
\end{figure}

\begin{figure}[htb]
\begin{center}
\epsfig{file=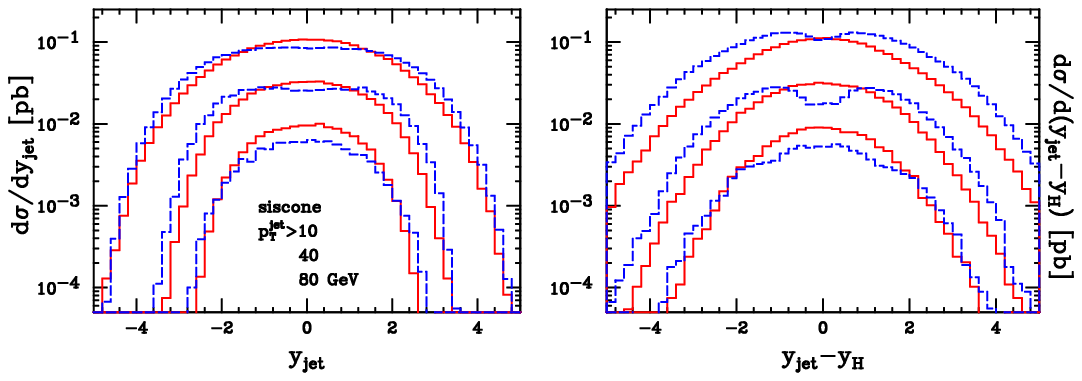,width=\figwidth}
\end{center}
\captskip
\caption{\label{fig:cmp2-tev-mcatnlo}
Comparison between \POWHEG{} and \MCatNLO{} for the rapidity of the leading
jet and the rapidity difference of the Higgs boson and the leading jet,
defined according to the \SISCONE{} algorithm, with different jet cuts.}
\end{figure}

\begin{figure}[htb]
\begin{center}
\epsfig{file=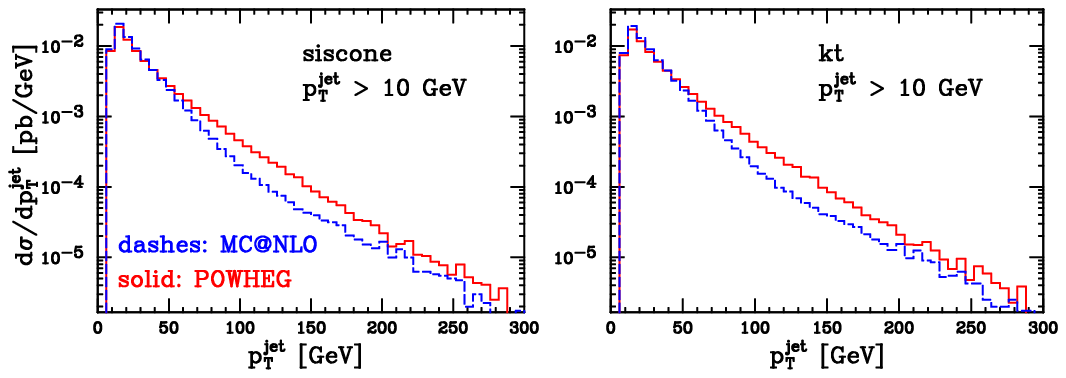,width=\figwidth}
\end{center}
\captskip
\caption{\label{fig:cmp3-tev-mcatnlo} Comparison between \POWHEG{} and
\MCatNLO{} for the transverse-momentum distributions of the leading jet,
defined according to the \SISCONE{} and the \FASTKT{} jet algorithms.}
\end{figure} 

In fig.~\ref{fig:cmp1-tev-mcatnlo} we show a comparison between \POWHEG\ and
\MCatNLO{} for the rapidity, invariant mass and transverse-momentum
distributions of a Higgs boson with mass $m_H =120$~GeV, at the Tevatron $p
\bar p$ collider. The lowest order $m_t$-dependence is retained.  A blowup of
the transverse-momentum distribution near the low-$\pt$ region is also
shown. There is 
good agreement between the two programs, except for the transverse momentum
distribution at high $\pt$ (we will comment more on this issue in
sec.~\ref{sec:hnnlo}).

In fig.~\ref{fig:cmp2-tev-mcatnlo} we compare the leading jet rapidity and
the difference in the rapidity of the leading jet and the Higgs boson. The
jet is defined using the \SISCONE{} algorithm~\cite{Salam:2007xv} as
implemented in the {\tt FASTJET} package~\cite{Cacciari:2005hq}, setting the
jet radius $R=0.7$ and the overlapping fraction $f=0.5$. As in previous
\POWHEG{} implementations, we notice a dip in the \MCatNLO{} jet rapidity
distribution, which is enhanced in the difference. We have already
extensively discussed this fact in sec.~4.3 of ref.~\cite{Alioli:2008gx}.

In fig.~\ref{fig:cmp3-tev-mcatnlo}, we compare the transverse-momentum
distributions of the leading jet, reconstructed with the \SISCONE{} and the
$k_T$ algorithms (included in {\tt FASTJET}).  A lower $10$~GeV cut on jet
transverse momentum is imposed. The high-$\pt$ discrepancy reflects the same
behavior found for the Higgs boson transverse-momentum distribution
(see sec.~\ref{sec:hnnlo}).

\subsubsection{LHC results}

\begin{figure}[htb]
\begin{center}
\epsfig{file=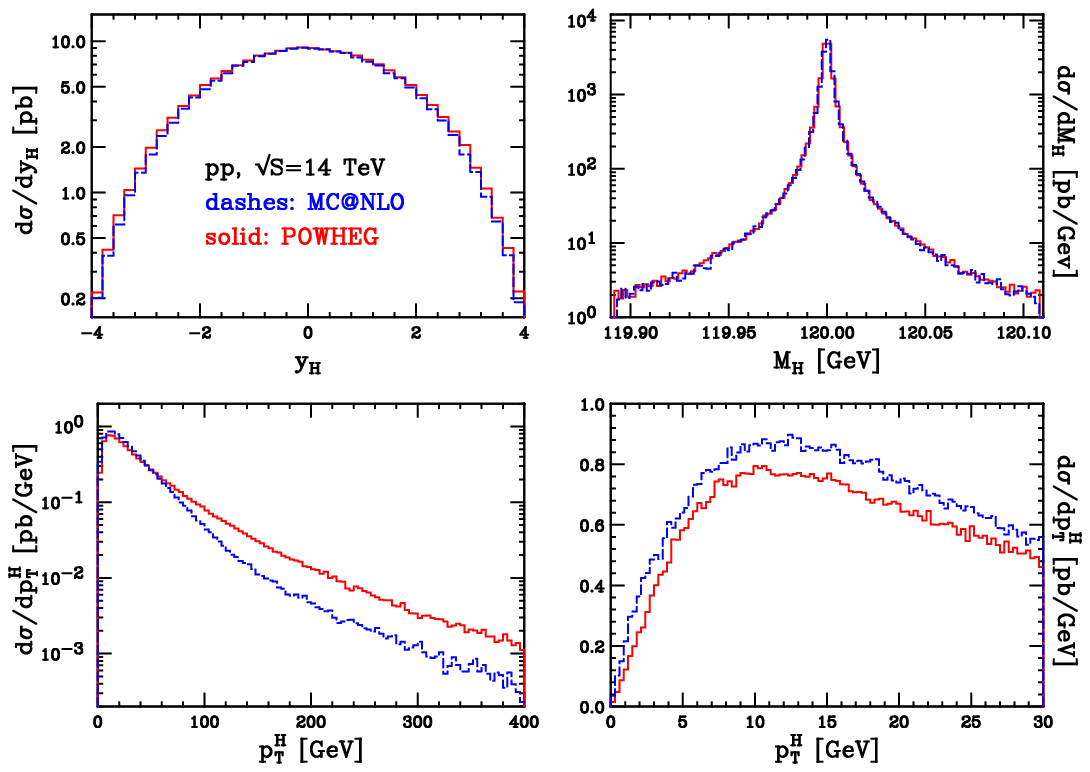,width=\figwidth}
\end{center}
\captskip
\caption{\label{fig:cmp1-lhc-mcatnlo}
Comparison between \POWHEG{} and \MCatNLO{} for the rapidity, invariant
mass and transverse-momentum distributions of a Higgs boson with $m_H
=120$~GeV, at the LHC $pp$ collider.} 
\end{figure}

\begin{figure}[htb]
\begin{center}
\epsfig{file=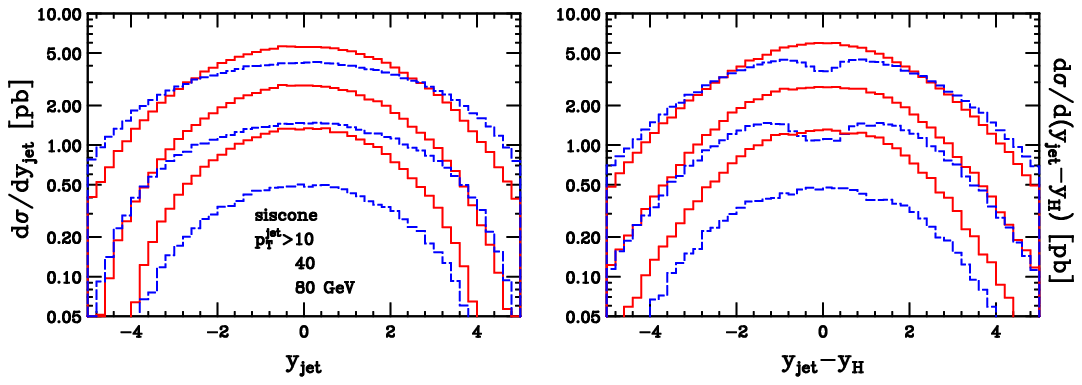,width=\figwidth}
\end{center}
\captskip
\caption{\label{fig:cmp2-lhc-mcatnlo}
Comparison between \POWHEG{} and \MCatNLO{} for the rapidity of the leading
jet and the rapidity difference of the Higgs boson and the leading jet,
defined according to the \SISCONE{} algorithm, with different jet
cuts.}
\end{figure}

\begin{figure}[htb]
\begin{center}
\epsfig{file=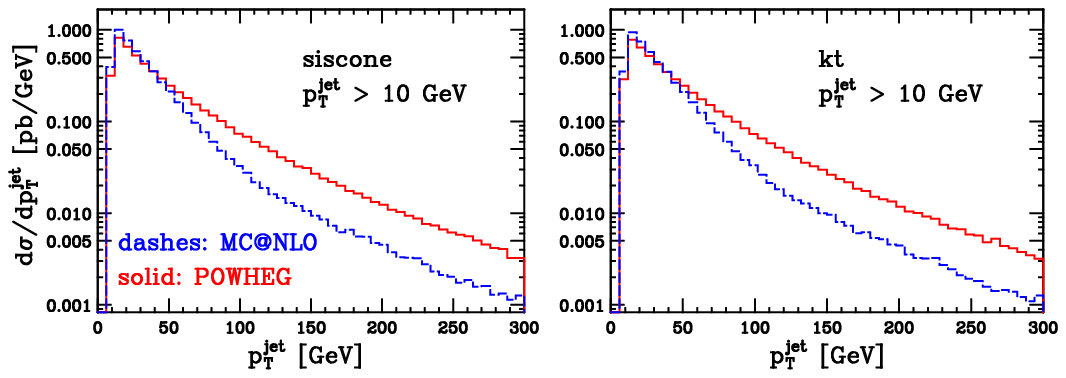,width=\figwidth}
\end{center}
\captskip
\caption{\label{fig:cmp3-lhc-mcatnlo} Comparison between \POWHEG{} and
\MCatNLO{} for the transverse-momentum distributions of the leading jet,
defined according to the \SISCONE{} and the \FASTKT{} algorithms.}
\end{figure}

\begin{figure}[htb]
\begin{center}
\epsfig{file=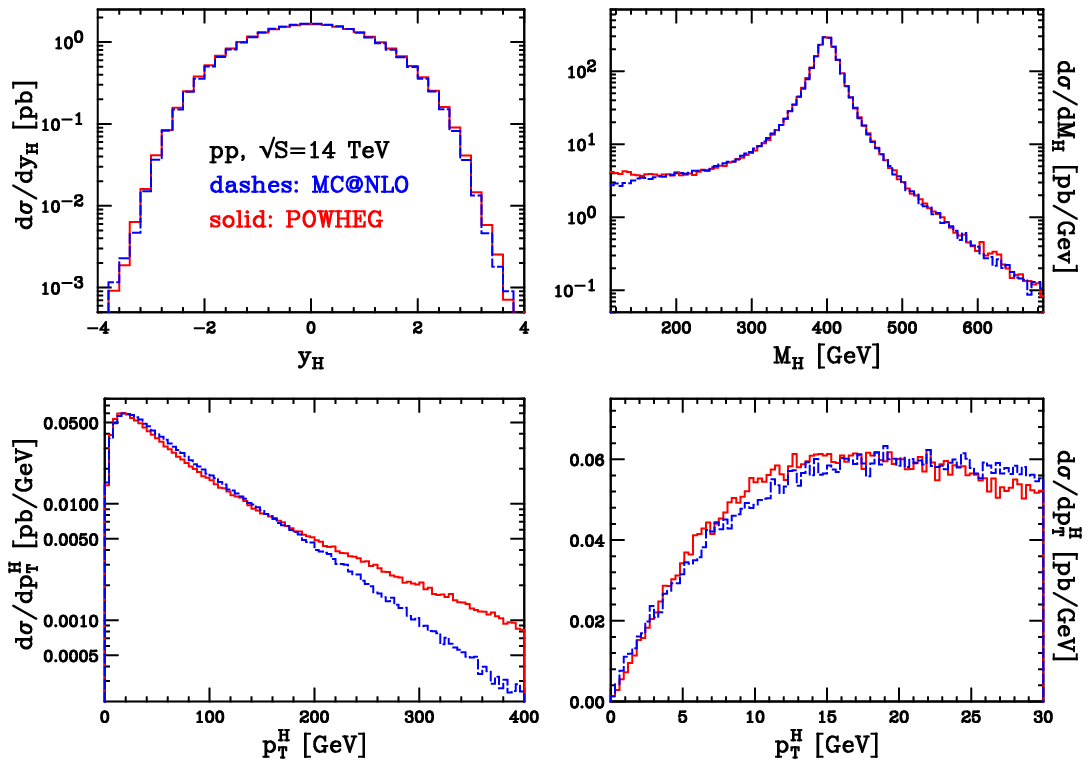,width=\figwidth}
\end{center}
\captskip
\caption{\label{fig:cmp1-lhc-mcatnlo-400}
Comparison between \POWHEG{} and \MCatNLO{} for the rapidity, invariant
mass and transverse-momentum distributions of a Higgs boson with $m_H
=400$~GeV, at the LHC $pp$ collider.} 
\end{figure}
%
                                                                          
\begin{figure}[htb]
\begin{center}
\epsfig{file=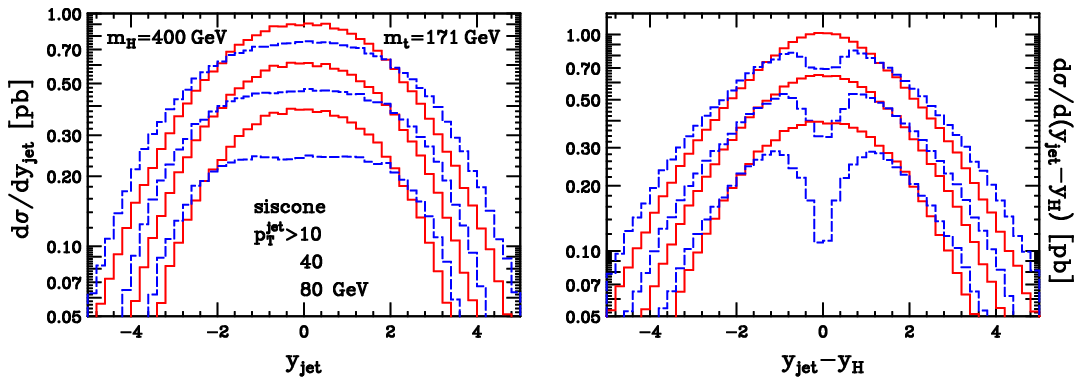,width=\figwidth}
\end{center}
\captskip
\caption{\label{fig:cmp2_400_LHC-POWHEG-MCatNLO}
Comparison between \POWHEG{} and \MCatNLO{} for the rapidity of the leading
jet and the rapidity difference of the Higgs boson and the leading jet,
defined according to the \SISCONE{} algorithm, with different jet
cuts.}  
\end{figure}

From fig.~\ref{fig:cmp1-lhc-mcatnlo} to~\ref{fig:cmp3-lhc-mcatnlo} we carry
out a similar analysis for the LHC $pp$ collider.  The difference in the
hardness of the $\pt$ distributions is more evident here than at the
Tevatron.  The other plots show instead a good agreement 
between the two codes, apart from the aforementioned dip in the leading-jet
rapidity distributions.

We have also made some comparisons with a different value of the Higgs boson
mass. We have chosen $m_H =400$~GeV, where
the ratio between the Born cross sections evaluated with $m_t=171$~GeV and
$m_t \to \infty$ is close to its maximum value and roughly equals 3.
The results are shown in fig.~\ref{fig:cmp1-lhc-mcatnlo-400}
and~\ref{fig:cmp2_400_LHC-POWHEG-MCatNLO}.
We see that, in this case, the dip in the rapidity of the hardest jet in
\MCatNLO{} is extremely marked.

\begin{figure}[htb]
\begin{center}
\epsfig{file=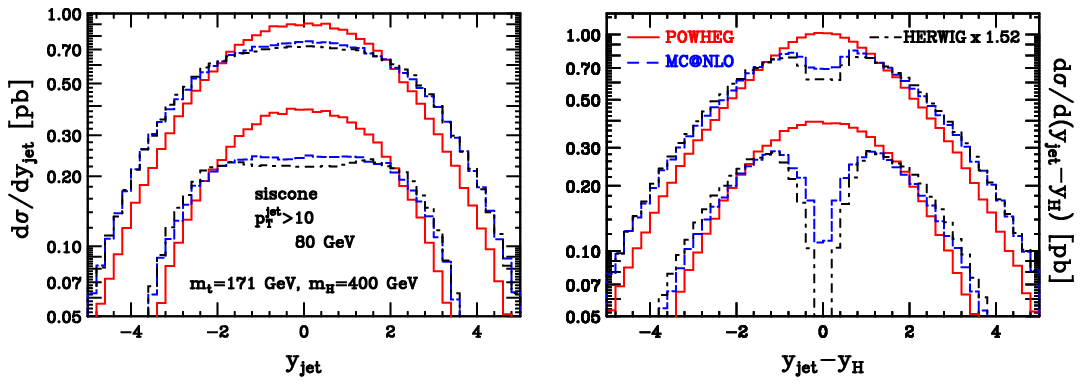,width=\figwidth}
\end{center}
\captskip
\caption{\label{fig:HERWIGdip}
  Comparison of \POWHEG{}, \MCatNLO{} and \HERWIG{} (without matrix-element
  corrections), for the rapidity of the 
  leading jet and the rapidity difference of the Higgs boson and the leading
  jet, defined according to the \SISCONE{} algorithm, with different jet
  cuts.}
\end{figure}

In the study of ref.~\cite{Mangano:2006rw}, carried out in the framework
of heavy-flavour production, the origin of the rapidity dip was tracked
back to an even stronger dip in the pure \HERWIG{} distribution,
that the \MCatNLO{} correction was not able to properly fill.
The same pattern is also observed in the present context, as can be seen
in fig.~\ref{fig:HERWIGdip}.

\subsection{\POWHEG\ - \PYTHIA\  comparison}
\begin{figure}[htb]
\begin{center}
\epsfig{file=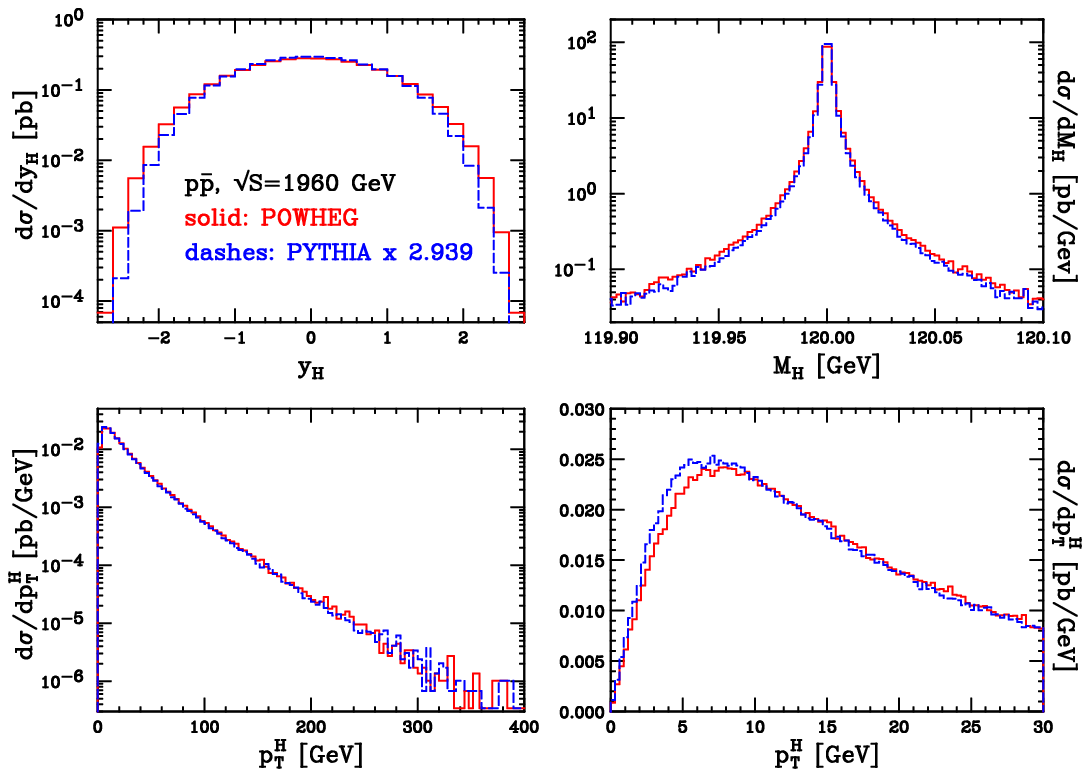,width=\figwidth}
\end{center}
\captskip
\caption{\label{fig:cmp1-tev-pythia}
Comparison between \POWHEG{} and \PYTHIA{} for the rapidity, invariant
mass and transverse-momentum distributions of a Higgs boson with $m_H
=120$~GeV, at Tevatron $p\bar p$ collider. \PYTHIA{} outputs normalized to the
\POWHEG{} cross section.} 
\end{figure}

\begin{figure}[htb]
\begin{center}
\epsfig{file=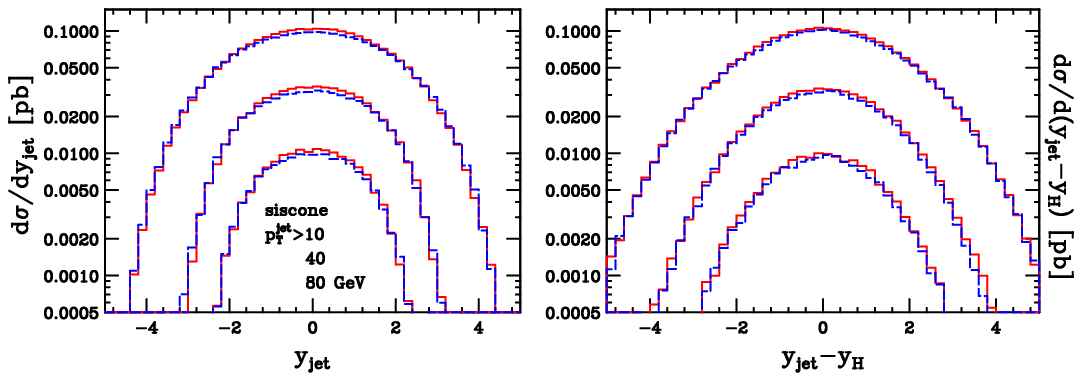,width=\figwidth}
\end{center}
\captskip
\caption{\label{fig:cmp2-tev-pythia}
Comparison between \POWHEG{} and \PYTHIA{} for the rapidity of the leading
jet and the rapidity difference of the Higgs boson and the leading jet,
defined according to the \SISCONE{} algorithm, with different jet
cuts. \PYTHIA{} outputs are normalized to the \POWHEG{} cross section.}
\end{figure}

\begin{figure}[htb]
\begin{center}
\epsfig{file=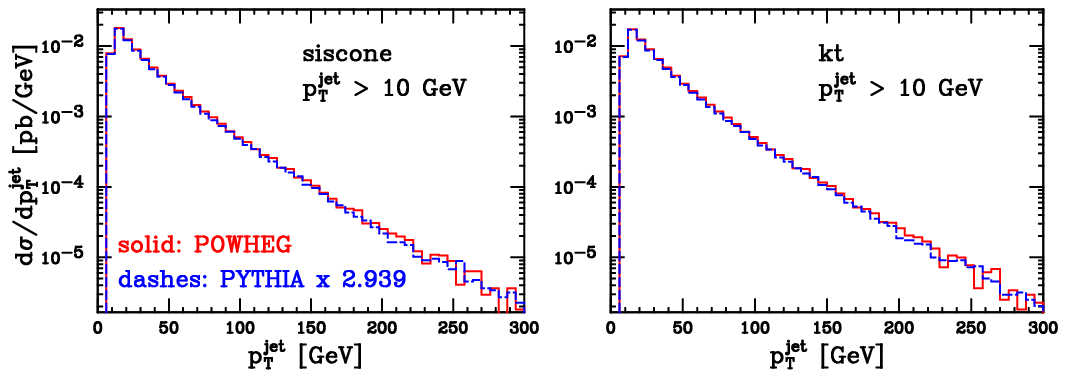,width=\figwidth}
\end{center}
\captskip
\caption{\label{fig:cmp3-tev-pythia} Comparison between \POWHEG{} and
\PYTHIA{} for the transverse-momentum distributions of the leading jet,
defined according to the \SISCONE{} and the \FASTKT{} algorithms. \PYTHIA{}
outputs are normalized to the
\POWHEG{} cross section.}
\end{figure}

\begin{figure}[htb]
\begin{center}
\epsfig{file=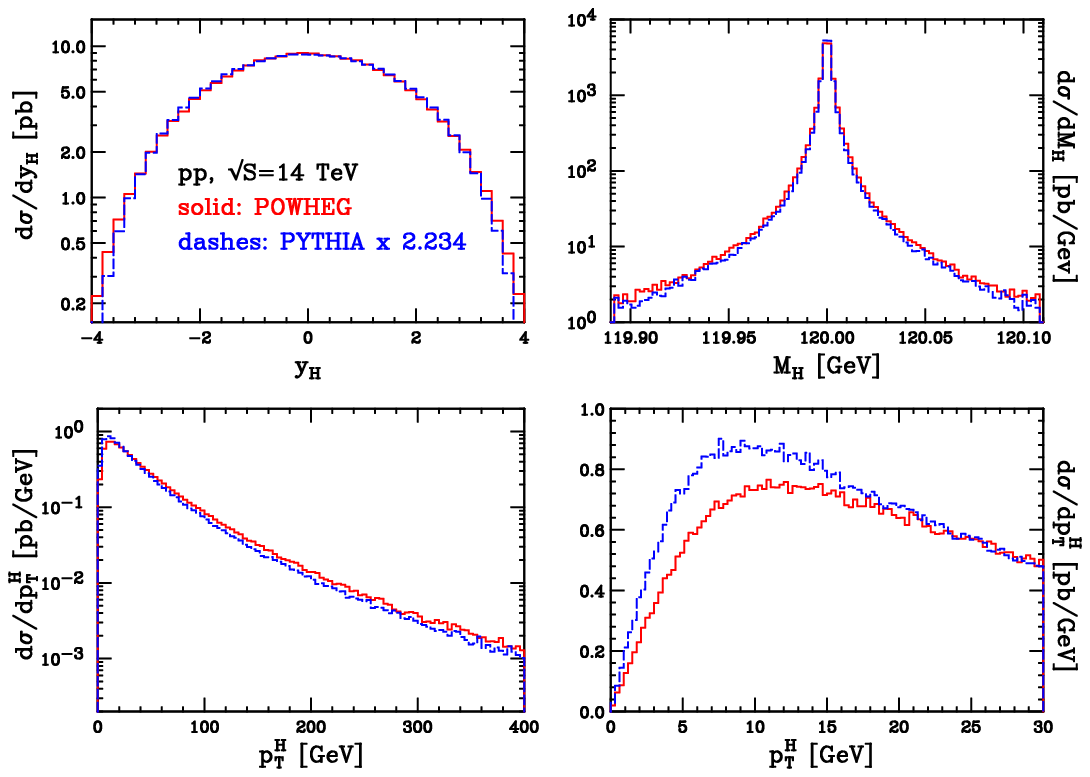,width=\figwidth}
\end{center}
\captskip
\caption{\label{fig:cmp1-lhc-pythia}
Comparison between \POWHEG{} and \PYTHIA{} for the rapidity, invariant
mass and transverse-momentum distributions of a Higgs boson with $m_H
=120$~GeV, at the LHC. \PYTHIA{} outputs are normalized to the 
\POWHEG{} cross section.} 
\end{figure}

\begin{figure}[htb]
\begin{center}
\epsfig{file=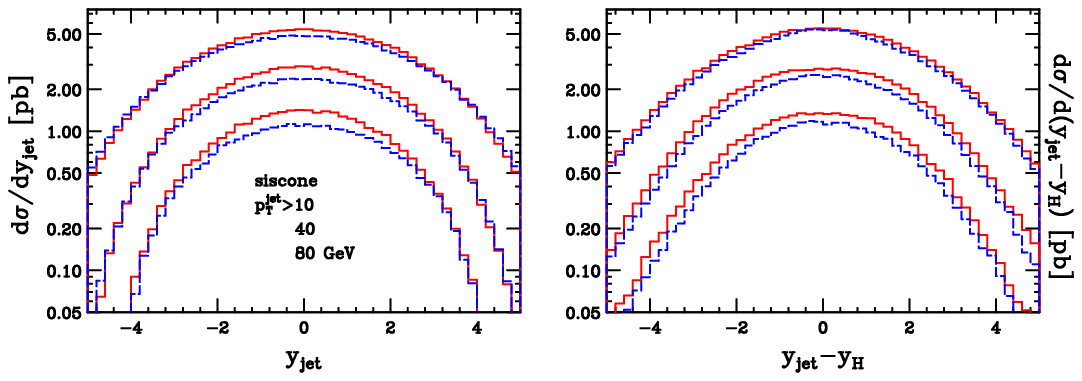,width=\figwidth}
\end{center}
\captskip
\caption{\label{fig:cmp2-lhc-pythia} Comparison between \POWHEG{} and
\PYTHIA{} for the rapidity of the leading jet and the rapidity
difference of the Higgs boson and the leading jet, defined according
to the \SISCONE{} algorithm, with different jet cuts. \PYTHIA{} outputs
are normalized to the
\POWHEG{} cross section.}
\end{figure}

\begin{figure}[htb]
\begin{center}
\epsfig{file=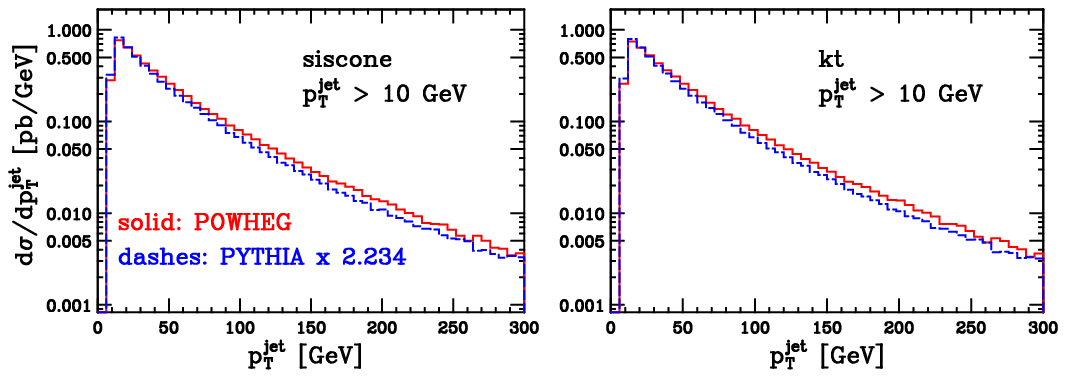,width=\figwidth}
\end{center}
\captskip
\caption{\label{fig:cmp3-lhc-pythia}
Comparison between \POWHEG{} and \PYTHIA{} for the transverse-momentum 
distributions of the leading jet, defined according to the \SISCONE{} and
the \FASTKT{} algorithms. \PYTHIA{} outputs are normalized to the
\POWHEG{} cross section.}
\end{figure}

\begin{figure}[htb]
\begin{center}
\epsfig{file=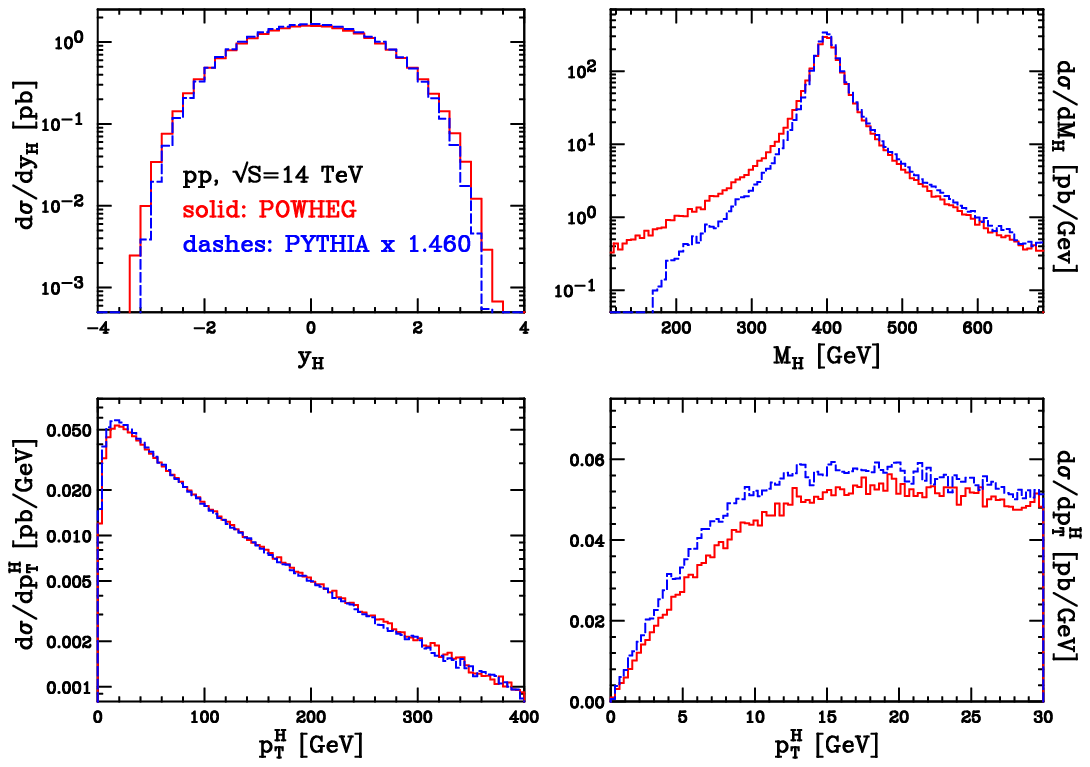,width=\figwidth}
\end{center}
\captskip
\caption{\label{fig:cmp1-lhc-pythia-400}
Comparison between \POWHEG{} and \PYTHIA{} for the rapidity, invariant
mass and transverse-momentum distributions of a Higgs boson with $m_H
=400$~GeV, at the LHC. \PYTHIA{} outputs are normalized to the 
\POWHEG{} cross section.} 
\end{figure}

We now compare \POWHEG{} and \PYTHIA{}. The Higgs boson
production implementation in \PYTHIA{} includes matrix-element
corrections, so that the $\pt$ distribution of the Higgs boson is
accurate at large $\pt$. In our comparisons, we always normalize
the \PYTHIA{} results to the full NLO cross section of \POWHEG{}.
We use the new $\pt$-ordered
shower defined in the \texttt{PYEVNW} routine of \PYTHIA, that should
be more appropriate when interfacing to \POWHEG{}.

The only difference with respect to the \POWHEG{}-\MCatNLO{}
comparisons is in the generation of the Higgs boson
virtuality, distributed now according to
\beq \frac{1}{\pi} \frac{M^2 \,\Gamma_H
  / m_H}{\(M^2 - m_H^2\)^2 + ( M^2\, \Gamma_H / m_H)^2 }\,,
\eeq 
which is very similar to the form used in \PYTHIA{}, except for the fact that
\PYTHIA{} includes threshold effects in the calculation of the Higgs boson
width. In fact, \PYTHIA{} uses a running $\Gamma_H(M^2)$, that increases when
a decay channel opens up.  The effects of using a fixed or a running
$\Gamma_H$ are more evident for a heavy Higgs boson, as will be shown in the
following.

In figs.~\ref{fig:cmp1-tev-pythia} through~\ref{fig:cmp3-tev-pythia} we
compare results for the Tevatron $p \bar p$ collider, while in
figs.~\ref{fig:cmp1-lhc-pythia} through~\ref{fig:cmp3-lhc-pythia} 
we present results for the LHC. In all the plots we have set 
$m_H=120$~GeV. Results are in an impressive good agreement, both for inclusive
quantities and for more exclusive ones.
The only visible difference is in
the transverse Higgs boson momentum distribution at low $\pt$ at the LHC.
This could be due to the different choice of the renormalization and
factorization scale in the generation of radiation, our choice being
constrained by the requirement of next-to-leading logarithmic accuracy in the
Sudakov form factor.

In fig.~\ref{fig:cmp1-lhc-pythia-400} we present a comparison with
$m_H=400$~GeV.  Mass thresholds effects in $\Gamma_H$ are evident in the
invariant-mass distribution generated by \PYTHIA{}.  Below $2\,m_t$ the total
width is smaller than the fixed one we are using, and \PYTHIA{} results are
accordingly lower than ours.  All other plots show instead good agreement
with \POWHEG.

The good agreement between \POWHEG{} and \PYTHIA{} was to some extent
expected.  As already observed in refs.~\cite{Frixione:2007vw,Alioli:2008gx},
the matrix-element correction method used in
\PYTHIA~\cite{Bengtsson:1986hr,Sjostrand:2006su} bears considerable
similarities to \POWHEG.

\subsection{The $\boldsymbol{\pt}$ distribution in \POWHEG{}}
\label{sec:hnnlo} 
\begin{figure}[htb]
\begin{center}
\epsfig{file=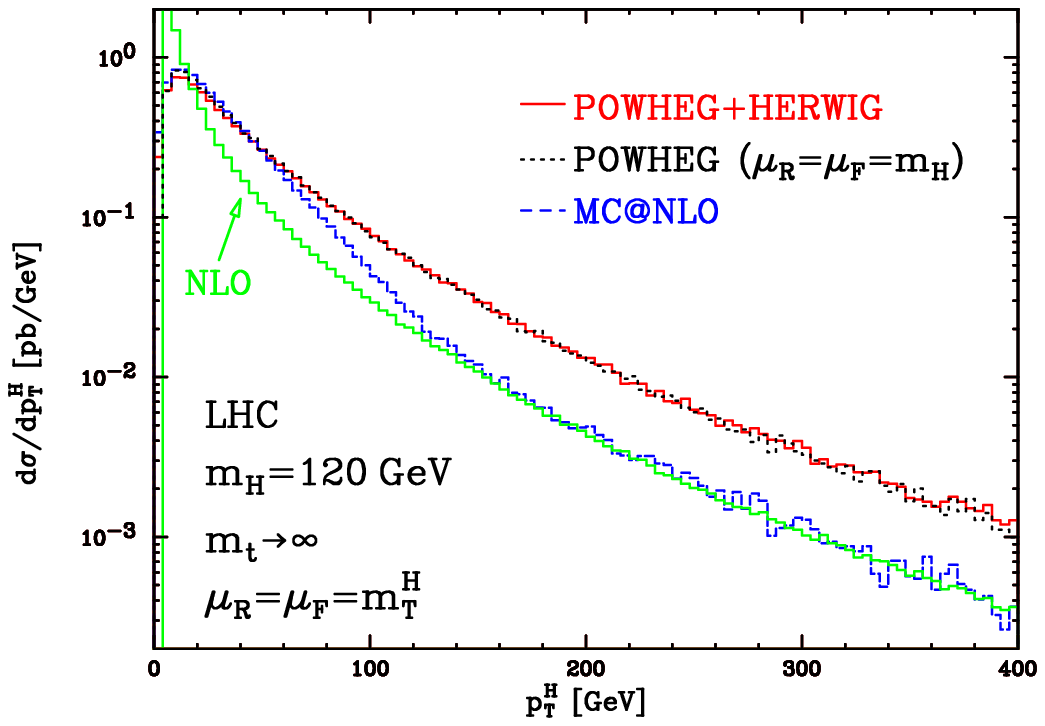,width=\figwidth}
\end{center}
\captskip
\caption{\label{fig:highptall}
Comparison between \POWHEG{}, \MCatNLO{} and the NLO calculation,
for $m_H=120$~GeV at the LHC.
All calculations are performed in the $m_t\to \infty$ approximation.
Shower and hadronization are included in the MC results.
The \POWHEG{} result is also presented without shower and
hadronization, and with a fixed-scale choice.
}

\end{figure}
\begin{figure}[htb]
\begin{center}
\epsfig{file=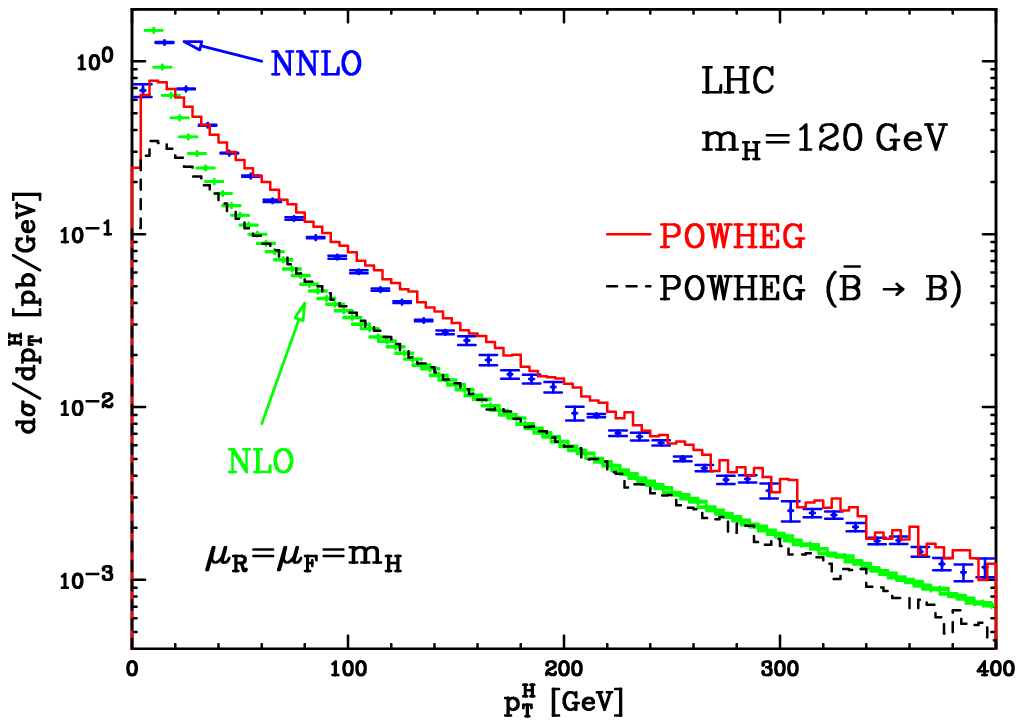,width=\figwidth}
\end{center}
\captskip
\caption{\label{fig:cmp-lhc-hnnlo} Comparison between \POWHEG{} and fixed NLO
and NNLO distributions for the transverse-momentum of the Higgs boson. Plots
are done for $m_H=120$~GeV at the LHC.}
\end{figure}

In this section we address the discrepancy in the $\pt$ distributions in
\POWHEG{} and in \MCatNLO. First of all, we show in fig.~\ref{fig:highptall}
a comparison between the $\pt$ spectrum of \POWHEG{}, \MCatNLO{} and the NLO
calculation.  For sake of comparison, we have used  in \POWHEG{} and in the NLO
calculation the same scale choice adopted in \MCatNLO. We point out,
however, that using a scale that depends upon the transverse momentum
of radiation in \POWHEG{} can only affect the $\bar{B}$ function. More
specifically, one ends up using a transverse momentum dependent scale
only in calculation of  the real contributions in $\bar{B}$,
since the transverse momentum is zero for
the Born, virtual and collinear remnant terms. Thus, this scale does not
depend upon the transverse momentum of the real radiation, that is generated
afterwards using the \POWHEG{} Sudakov form factor. The choice of scale
for radiation affects instead a single power of the coupling constant,
since the Sudakov exponent is proportional to $\as$.
At low transverse momentum, this scale
cannot be changed without spoiling the NLL accuracy of the
Sudakov form factor. It can be changed, however, at large transverse momentum
to explore further uncertainties. However, we have preferred
not to implement this possibility.
One should recall, in fact, that this
scale only affects a single power of $\as$, and it thus has a much
smaller effect than a scale change in the NLO cross section.

We see from fig.~\ref{fig:highptall} that \MCatNLO{} agrees better
than \POWHEG{}
with the NLO calculation at large $\pt$. Since the difference between
\MCatNLO{} and \POWHEG{} should be of next-to-next-to-leading order (NNLO),
the difference between \POWHEG{} and the NLO result should also be
of NNLO. In fact we can easily trace the origin of this difference.
From eq.~(\ref{eq:POWHEGsigmasimple}), we infer that, at large $\pt$,
the \POWHEG{} differential cross section can be written as
\begin{equation}
\label{eq:POWHEGsigmasimple1}
d\sigma= \lq \bar{B}(\bar{\bf \Phi}_1)\, 
\frac{ R\({\bf \Phi}_2\)}{ B\!\(\bar{\bf \Phi}_1\)}  + \sum_q R_{q \bar q}\({\bf
  \Phi}_2\) \rq \, d \bar{\bf
  \Phi}_1 \, d\Rad\,, 
\end{equation}
since the Sudakov form factor approaches 1 in this region. Neglecting the
subdominant $q\bar{q}$ real contribution, this differs
from the pure NLO result because of the presence of the factor
\begin{equation}
\frac{\bar{B}(\bar{\bf \Phi}_1)}{B\!\(\bar{\bf \Phi}_1\)}=1+{\cal O}(\as)\;.
\end{equation}
It is known that radiative corrections in Higgs boson production are large, so
that the ${\cal O}(\as)$ term is in fact of order 1, and thus we find an
enhancement that approaches a factor of two.\footnote{We recall that normally
the numerator and denominator in this factor are evaluated at different
scales, since in $\bar{B}$ one uses a scale of the order of the Higgs boson
transverse mass, while in the $B$ term, one uses the transverse momentum. However,
at large $\pt$, the two scales become of the same order.}  We have performed
a clear cut test of this interpretation of the discrepancy. We have replaced
the $\bar{B}$ function with the Born term $B$ in the \POWHEG{} program. The
result of this calculation is shown in comparison with the NLO curve in
fig.~\ref{fig:cmp-lhc-hnnlo}.  Since, as shown in fig.~\ref{fig:highptall},
the shower and hadronization are irrelevant for this distribution, we do not
include them in the figure. In fig.~\ref{fig:cmp-lhc-hnnlo} we have chosen
to use $\pt$ independent renormalization and factorization scales, in order
to perform a consistent comparison. Notice that, with this choice of scales,
the NLO distribution is harder than the one shown in fig.~\ref{fig:highptall}.
This is easily explained by the fact that the NLO process is proportional
to $\as^3(\mur)$, and thus a $\pt$ dependent renormalization scale can
alter significantly the $\pt$ distribution.

At this point, we can ask whether the higher
order terms included in \POWHEG{} with the mechanism illustrated above do in
fact give a reasonable estimate of true NNLO effects.  We thus include in
fig.~\ref{fig:cmp-lhc-hnnlo} the NNLO result, obtained from the \texttt{HNNLO}
program of ref.~\cite{Catani:2007vq}.  The result shows a rather good
agreement between the NNLO result and \POWHEG{}. Thus, our seemingly large
corrections to the Higgs boson $\pt$ distributions are in fact very similar
in size to the full NNLO result. Observe that in
fig.~\ref{fig:cmp-lhc-hnnlo} we have used a fixed scale choice for all the
results. We were forced to do this, since the \texttt{HNNLO} program does not
allow for other choices. However, because of the good agreement of the two
\POWHEG{} results in fig.~\ref{fig:highptall}, and because of the smaller
scale dependence of the NNLO result, this should not make a severe
difference.

\begin{figure}[htb]
\begin{center}
\epsfig{file=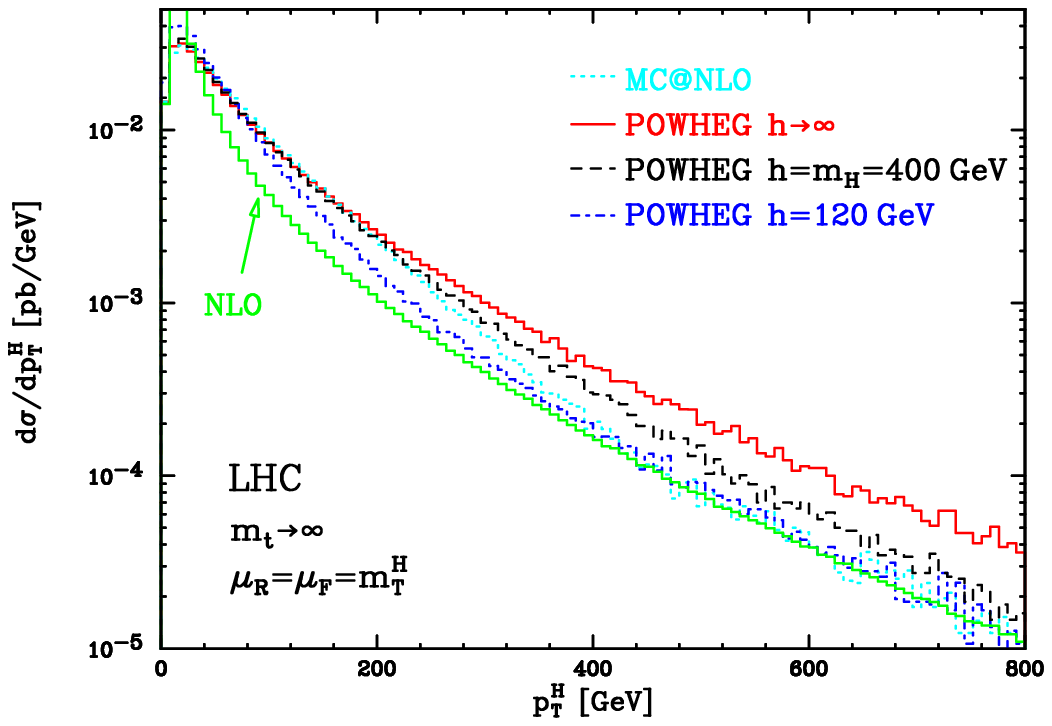,width=\figwidth}
\end{center}
\captskip
\caption{\label{fig:ptwithF} Comparison of the predictions of \MCatNLO{},
  standard \POWHEG{} ($h\to\infty$) and \POWHEG{} with two different values of
  the parameter $h$ ($h=120$~GeV and $h=m_H=400$~GeV) in the function $F$ of
  eq.~(\ref{eq:freduction}), for the transverse-momentum distributions of a
  Higgs boson, at the LHC $pp$ collider.}
\end{figure}

\begin{figure}[htb]
\begin{center}
\epsfig{file=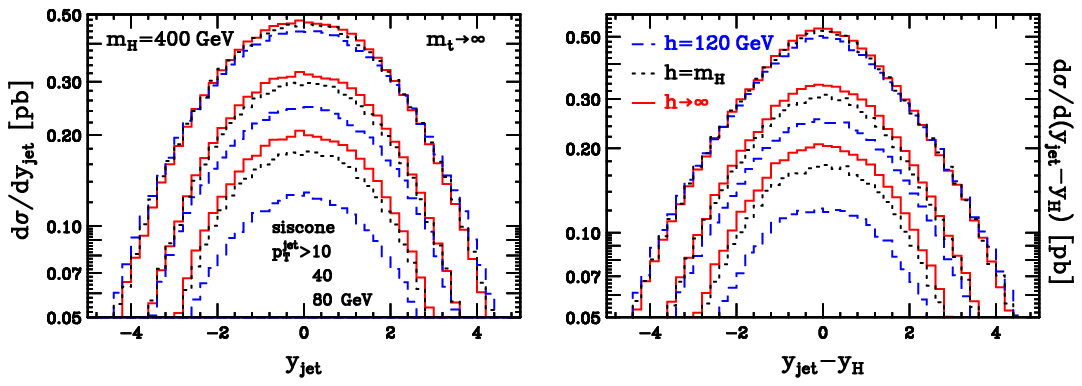,width=\figwidth}
\end{center}
\captskip
\caption{\label{fig:rapwithF} Comparison of the predictions of the standard
  \POWHEG{} ($h\to\infty$), and \POWHEG{} with two different values of the
  parameter $h$ ($h=120$~GeV and $h=m_H=400$~GeV) in the function $F$ of
  eq.~(\ref{eq:freduction}), for the rapidity of the leading jet and the
  rapidity difference of the Higgs boson and the leading jet, defined
  according to the \SISCONE{} algorithm, with different jet cuts, at the LHC.}
\end{figure}

Because of a fortuitous circumstance, we did not need to worry about
correcting for the large difference between the \POWHEG{} and the NLO
result at large radiation transverse momentum, since the known NNLO
result seems to support the \POWHEG{} one. We remark, however,
that, had this not been the case, it is very easy to modify
the \POWHEG{} algorithm so to obtain a $\pt$ spectrum that agrees
with the NLO calculation at large $\pt$. This can be done as follows.
Instead of using the full real cross section for the computation of the
$\bar{B}$ function and of the Sudakov form factor, we can instead
use a reduced real contribution
\begin{equation}
R^{\rm red}=R \times F\,,
\end{equation}
where $F$ is a function of the real phase space, with $F<1$ everywhere, such
that $F$ approaches 1 for small transverse momenta, and approaches zero for
large transverse momenta. We perform the \POWHEG{} generation using $R^{\rm
red}$ instead of $R$, and treat the remaining $R\times(1-F)$ contribution to
the cross section with the same method that we used for the $R_{q\bar{q}}$
contribution. This can be done, since $R\times(1-F)$ is dumped by the $1-F$
factor in the singular region. It will then follow that, for large transverse
momentum, the result would agree with the NLO calculation, since it would be
dominated by the $R\times(1-F)$ contribution.  It turned out that, in all
previous implementations, it was not necessary to use such procedure. As
remarked before, thanks to the known properties of the NNLO result, this was
not necessary even in this case.  
We have however performed such study, just in order to illustrate the
flexibility of the \POWHEG{} method. We have chosen for $F$ the following
form
\begin{equation}
\label{eq:freduction}
F=\frac{h^2}{\pt^2+h^2}\,.
\end{equation}
The resulting transverse-momentum distribution at the LHC, for a Higgs boson
mass of 400~GeV, is shown in fig.~\ref{fig:ptwithF} for $h\to\infty$ (standard
\POWHEG{}), $h=120$~GeV and $h=400$~GeV.
One can see that it is not difficult to get distributions that undershoot the
\MCatNLO{} one in the intermediate range of $\pt$.  We also observe that, with
this procedure, no undesired features of other distributions appear. In
particular, the distribution in the rapidity of the hardest jet, and in the
rapidity difference between the hardest jet and the Higgs boson remain
qualitatively the same, as shown in fig.~\ref{fig:rapwithF}.

\subsection{Next-to-leading logarithmic resummation}
\label{sec:nllresum}
\begin{figure}[htb]
\begin{center}
\epsfig{file=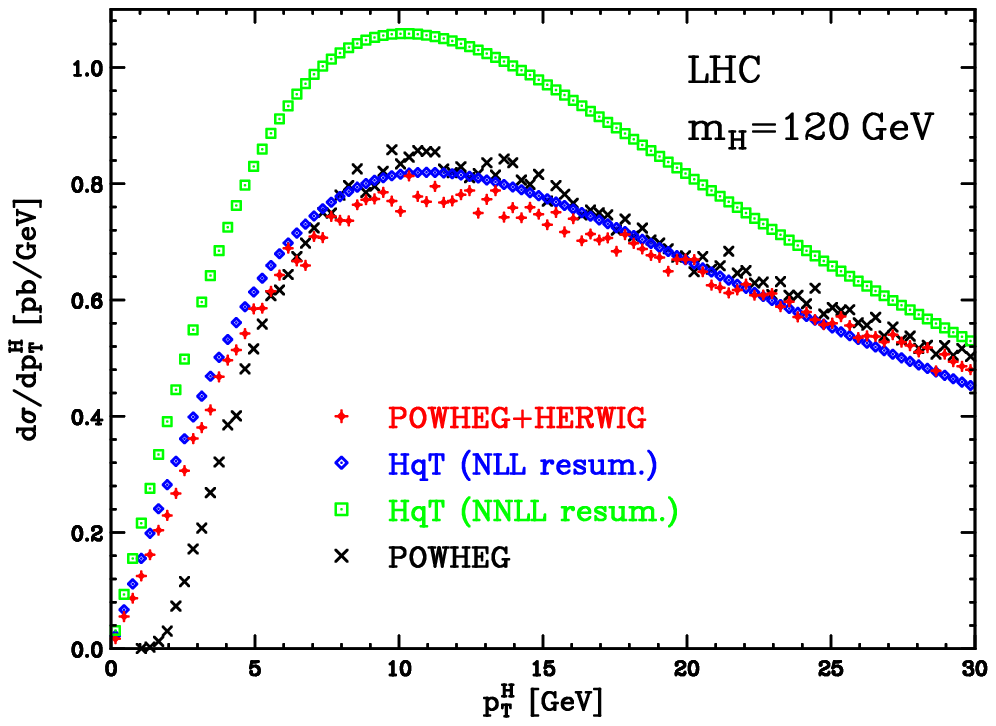,width=\figwidth}
\end{center}
\captskip
\caption{\label{fig:cmp-lhc-hqt} Comparison between \POWHEG{} and
\texttt{HqT} for the transverse-momentum distributions of a Higgs
boson, at the LHC.  The \POWHEG{} result without shower and hadronization is also
shown.} 
\end{figure}
As explained in section~(4.4) of ref.~\cite{Frixione:2007vw}, one can reach
next-to-leading logarithmic (NLL) accuracy of soft gluon resummation if the
number of coloured partons involved in the hard scattering is less or
equal to three. This can be obtained by replacing the strong coupling
constant in the Sudakov exponent with~\cite{Catani:1991rr}
\begin{equation}
    \label{eq:nllcond} \as \rightarrow A \left( \as \left( \kt^2
    \right) \right), \qquad A (\as) = \as \lg 1 + \frac{\as}{2 \pi}
    \lq \left( \frac{67}{18} - \frac{\pi^2}{6} \right) \CA - \frac{5}{9}
    \NF \rq \rg,
\end{equation}
where the $\overline{\tmop{MS}}$, 1-loop expression of $\as$ should be used.
The previous replacement may also be implemented by a simple redefinition of
the strong scale $\Lambda$, which, for five active flavours ($\NF=5$), becomes
$\Lambda_{\scriptscriptstyle\rm{MC}} \equiv 1.569\,
\Lambda_{\scriptscriptstyle\overline{\rm MS}}^{(5)}$.  We have exploited this
possibility in our code, so that our result should agree with the NLL
resummed one. A comparison has been thus carried out with the
\texttt{HqT}~\cite{Bozzi:2005wk} program, that performs such a resummation.
We have adopted fixed renormalization and factorization scales.  Results are
shown in fig.~\ref{fig:cmp-lhc-hqt}, together with next-to-next-to-leading
logarithmic (NNLL) resummation,
always from \texttt{HqT}, just for reference purposes.  We see a fair
agreement between the \POWHEG{} result and the NLL analytic one, as
expected. The different behaviour of the \POWHEG{} result without shower and
hadronization at very low $\pt$ may be ascribed to the particular
implementation of the minimum transverse momentum that we use, that is, to a
large extent, arbitrary.

We observe that, in all cases, we do not expect full agreement between the
\POWHEG{} result without shower, and the NLL calculation. In fact, the
\POWHEG{} curve without shower represents the Sudakov form factor for the
$\pt$ of the hardest emission, while, in the NLL calculation, the 
total $\pt$ distribution (i.e.~the sum of the transverse momenta of all
emissions) is considered. Thus, it is only after the inclusion of the full
shower effects that the two distributions have a meaningful comparison.

\section{Conclusions}
\label{sec:conc}
In this paper we have reported on a complete implementation of Higgs boson
production via gluon fusion at next-to-leading order in QCD, in the \POWHEG{}
framework. The calculation was performed within the
Frixione-Kunszt-Signer~\cite{Frixione:1995ms,Frixione:1997np} subtraction
approach. We have also shown how to deal with non-singular real
contributions, that do not present a valid underlying Born matrix element.

The results of our work have been compared extensively with \MCatNLO{} and
\PYTHIA{} shower Monte Carlo programs. The \PYTHIA{} results, normalized to
the total NLO cross section, are in good agreement with \POWHEG{}, except for
differences in the low transverse-momentum distributions of the Higgs boson
at the LHC. The \MCatNLO{} results are in fair agreement with \POWHEG, except
for the $\pt$ distribution of the Higgs boson, and consequently of the
hardest jet, in the high-$\pt$ region. In this region the \POWHEG{}
distributions are generally harder.  We have shown that this is due to NNLO
effects in the \POWHEG{} formula for the differential cross section.  We
checked that these effects actually bring our result closer to the NNLO
one~\cite{Catani:2007vq}. The low-$\pt$ region was instead tested against the
analytic resummed results~\cite{Bozzi:2005wk}. We find again
good agreement up to NLL accuracy.

Furthermore, we have also examined the distributions in the difference of the
hardest jet and the Higgs boson rapidity. The dip found in previous
implementations~\cite{Nason:2006hfa,Frixione:2007nw,Alioli:2008gx} is still
present.  We remark that this seems to be a general feature of \MCatNLO{},
since other calculations do not find effects of this
kind~\cite{Mangano:2006rw,Alwall:2007fs,Dittmaier:2008uj}.

The computer code for the \POWHEG{} implementations presented in this paper
is available, together with the manual, at the site\\ \\
\centerline{{\tt http://moby.mib.infn.it/\~{}nason/POWHEG\ }}


\bibliography{paper}

\providecommand{\href}[2]{#2}\begingroup\raggedright\begin{thebibliography}{10}

\bibitem{Dawson:1990zj}
S.~Dawson, {\it {Radiative corrections to Higgs boson production}},  {\em Nucl.
  Phys.} {\bf B359} (1991) 283--300.

\bibitem{Djouadi:1991tka}
A.~Djouadi, M.~Spira, and P.~M. Zerwas, {\it {Production of Higgs bosons in
  proton colliders: QCD corrections}},  {\em Phys. Lett.} {\bf B264} (1991)
  440--446.

\bibitem{Spira:1995rr}
M.~Spira, A.~Djouadi, D.~Graudenz, and P.~M. Zerwas, {\it {Higgs boson
  production at the LHC}},  {\em Nucl. Phys.} {\bf B453} (1995) 17--82,
  [\href{http://xxx.lanl.gov/abs/hep-ph/9504378}{{\tt hep-ph/9504378}}].

\bibitem{Frixione:2002ik}
S.~Frixione and B.~R. Webber, {\it {Matching NLO QCD computations and parton
  shower simulations}},  {\em JHEP} {\bf 06} (2002) 029,
  [\href{http://xxx.lanl.gov/abs/hep-ph/0204244}{{\tt hep-ph/0204244}}].

\bibitem{Corcella:2000bw}
G.~Corcella {\em et~al.}, {\it {HERWIG 6: An event generator for hadron
  emission reactions with interfering gluons (including supersymmetric
  processes)}},  {\em JHEP} {\bf 01} (2001) 010,
  [\href{http://xxx.lanl.gov/abs/hep-ph/0011363}{{\tt hep-ph/0011363}}].

\bibitem{Corcella:2002jc}
G.~Corcella {\em et~al.}, {\it Herwig 6.5 release note},
  \href{http://xxx.lanl.gov/abs/hep-ph/0210213}{{\tt hep-ph/0210213}}.

\bibitem{Sjostrand:2006za}
T.~Sjostrand, S.~Mrenna, and P.~Skands, {\it Pythia 6.4 physics and manual},
  {\em JHEP} {\bf 05} (2006) 026,
  [\href{http://xxx.lanl.gov/abs/hep-ph/0603175}{{\tt hep-ph/0603175}}].

\bibitem{Nason:2004rx}
P.~Nason, {\it {A new method for combining NLO QCD with shower Monte Carlo
  algorithms}},  {\em JHEP} {\bf 11} (2004) 040,
  [\href{http://xxx.lanl.gov/abs/hep-ph/0409146}{{\tt hep-ph/0409146}}].

\bibitem{Frixione:2007vw}
S.~Frixione, P.~Nason, and C.~Oleari, {\it {Matching NLO QCD computations with
  Parton Shower simulations: the POWHEG method}},  {\em JHEP} {\bf 11} (2007)
  070, [\href{http://xxx.lanl.gov/abs/arXiv:0709.2092 [hep-ph]}{{\tt
  arXiv:0709.2092 [hep-ph]}}].

\bibitem{Nason:2006hfa}
P.~Nason and G.~Ridolfi, {\it {A positive-weight next-to-leading-order Monte
  Carlo for $Z$ pair hadroproduction}},  {\em JHEP} {\bf 08} (2006) 077,
  [\href{http://xxx.lanl.gov/abs/hep-ph/0606275}{{\tt hep-ph/0606275}}].

\bibitem{Frixione:2007nw}
S.~Frixione, P.~Nason, and G.~Ridolfi, {\it {A Positive-Weight
  Next-to-Leading-Order Monte Carlo for Heavy Flavour Hadroproduction}},  {\em
  JHEP} {\bf 09} (2007) 126, [\href{http://xxx.lanl.gov/abs/arXiv:0707.3088
  [hep-ph]}{{\tt arXiv:0707.3088 [hep-ph]}}].

\bibitem{LatundeDada:2006gx}
O.~Latunde-Dada, S.~Gieseke, and B.~Webber, {\it {A positive-weight
  next-to-leading-order Monte Carlo for $e^+ e^-$ annihilation to hadrons}},
  {\em JHEP} {\bf 02} (2007) 051,
  [\href{http://xxx.lanl.gov/abs/hep-ph/0612281}{{\tt hep-ph/0612281}}].

\bibitem{LatundeDada:2008bv}
O.~Latunde-Dada, {\it {Applying the POWHEG method to top pair production and
  decays at the ILC}},  \href{http://xxx.lanl.gov/abs/0806.4560}{{\tt
  0806.4560}}.

\bibitem{Alioli:2008gx}
S.~Alioli, P.~Nason, C.~Oleari, and E.~Re, {\it {NLO vector-boson production
  matched with shower in POWHEG}},  {\em JHEP} {\bf 07} (2008) 060,
  [\href{http://xxx.lanl.gov/abs/0805.4802}{{\tt 0805.4802}}].

\bibitem{Hamilton:2008pd}
K.~Hamilton, P.~Richardson, and J.~Tully, {\it {A Positive-Weight
  Next-to-Leading Order Monte Carlo Simulation of Drell-Yan Vector Boson
  Production}},  {\em JHEP} {\bf 10} (2008) 015,
  [\href{http://xxx.lanl.gov/abs/0806.0290}{{\tt 0806.0290}}].

\bibitem{Catani:2007vq}
S.~Catani and M.~Grazzini, {\it {An NNLO subtraction formalism in hadron
  collisions and its application to Higgs boson production at the LHC}},  {\em
  Phys. Rev. Lett.} {\bf 98} (2007) 222002,
  [\href{http://xxx.lanl.gov/abs/hep-ph/0703012}{{\tt hep-ph/0703012}}].

\bibitem{Mangano:2006rw}
M.~L. Mangano, M.~Moretti, F.~Piccinini, and M.~Treccani, {\it Matching matrix
  elements and shower evolution for top-quark production in hadronic
  collisions},  {\em JHEP} {\bf 01} (2007) 013,
  [\href{http://xxx.lanl.gov/abs/hep-ph/0611129}{{\tt hep-ph/0611129}}].

\bibitem{Alwall:2007fs}
J.~Alwall {\em et~al.}, {\it {Comparative study of various algorithms for the
  merging of parton showers and matrix elements in hadronic collisions}},  {\em
  Eur. Phys. J.} {\bf C53} (2008) 473--500,
  [\href{http://xxx.lanl.gov/abs/arXiv:0706.2569 [hep-ph]}{{\tt arXiv:0706.2569
  [hep-ph]}}].

\bibitem{Dittmaier:2008uj}
S.~Dittmaier, P.~Uwer, and S.~Weinzierl, {\it {Hadronic top-quark pair
  production in association with a hard jet at next-to-leading order QCD:
  Phenomenological studies for the Tevatron and the LHC}},
  \href{http://xxx.lanl.gov/abs/arXiv:0810.0452 [hep-ph]}{{\tt arXiv:0810.0452
  [hep-ph]}}.

\bibitem{Frixione:1995ms}
S.~Frixione, Z.~Kunszt, and A.~Signer, {\it {Three-jet cross sections to
  next-to-leading order}},  {\em Nucl. Phys.} {\bf B467} (1996) 399--442,
  [\href{http://xxx.lanl.gov/abs/hep-ph/9512328}{{\tt hep-ph/9512328}}].

\bibitem{Frixione:1997np}
S.~Frixione, {\it {A general approach to jet cross sections in QCD}},  {\em
  Nucl. Phys.} {\bf B507} (1997) 295--314,
  [\href{http://xxx.lanl.gov/abs/hep-ph/9706545}{{\tt hep-ph/9706545}}].

\bibitem{Nason:2007vt}
P.~Nason, {\it {MINT: a Computer Program for Adaptive Monte Carlo Integration
  and Generation of Unweighted Distributions}},
  \href{http://xxx.lanl.gov/abs/arXiv:0709.2085 [hep-ph]}{{\tt arXiv:0709.2085
  [hep-ph]}}.

\bibitem{Nason:2006hf}
P.~Nason and G.~Ridolfi, {\it {A positive-weight next-to-leading-order Monte
  Carlo for $Z$ pair hadroproduction}},  {\em JHEP} {\bf 08} (2006) 077,
  [\href{http://xxx.lanl.gov/abs/hep-ph/0606275}{{\tt hep-ph/0606275}}].

\bibitem{Pumplin:2002vw}
J.~Pumplin {\em et~al.}, {\it {New generation of parton distributions with
  uncertainties from global QCD analysis}},  {\em JHEP} {\bf 07} (2002) 012,
  [\href{http://xxx.lanl.gov/abs/hep-ph/0201195}{{\tt hep-ph/0201195}}].

\bibitem{Salam:2007xv}
G.~P. Salam and G.~Soyez, {\it {A practical Seedless Infrared-Safe Cone jet
  algorithm}},  {\em JHEP} {\bf 05} (2007) 086,
  [\href{http://xxx.lanl.gov/abs/arXiv:0704.0292 [hep-ph]}{{\tt arXiv:0704.0292
  [hep-ph]}}].

\bibitem{Cacciari:2005hq}
M.~Cacciari and G.~P. Salam, {\it {Dispelling the $N^3$ myth for the $k_T$
  jet-finder}},  {\em Phys. Lett.} {\bf B641} (2006) 57--61,
  [\href{http://xxx.lanl.gov/abs/hep-ph/0512210}{{\tt hep-ph/0512210}}].

\bibitem{Bengtsson:1986hr}
M.~Bengtsson and T.~Sjostrand, {\it {Coherent Parton Showers Versus Matrix
  Elements: Implications of PETRA - PEP Data}},  {\em Phys. Lett.} {\bf B185}
  (1987) 435.

\bibitem{Sjostrand:2006su}
T.~Sjostrand, {\it {Monte Carlo generators}},
  \href{http://xxx.lanl.gov/abs/hep-ph/0611247}{{\tt hep-ph/0611247}}.

\bibitem{Catani:1991rr}
S.~Catani, B.~R. Webber, and G.~Marchesini, {\it {QCD coherent branching and
  semiinclusive processes at large $x$}},  {\em Nucl. Phys.} {\bf B349} (1991)
  635--654.

\bibitem{Bozzi:2005wk}
G.~Bozzi, S.~Catani, D.~de~Florian, and M.~Grazzini, {\it {Transverse-momentum
  resummation and the spectrum of the Higgs boson at the LHC}},  {\em Nucl.
  Phys.} {\bf B737} (2006) 73--120,
  [\href{http://xxx.lanl.gov/abs/hep-ph/0508068}{{\tt hep-ph/0508068}}].

\end{thebibliography}\endgroup

\end{document}